\documentclass[aps,prd,twocolumn,superscriptaddress,nofootinbib,floatfix,longbibliography,letterpaper]{revtex4-2}

\usepackage[T1]{fontenc}
\usepackage{XCharter}
\usepackage[cmbraces]{newtxmath}
\usepackage{graphicx}
\usepackage[dvipsnames]{xcolor}
\usepackage{booktabs,array,multirow}
\usepackage{tikz}
\usepackage[compat=1.1.0]{tikz-feynman}
\usepackage{hyperref}
\hypersetup{colorlinks=true,linkcolor=NavyBlue,citecolor=NavyBlue,urlcolor=NavyBlue}
\usepackage{shorthand}
\definecolor{bsmred}{HTML}{D32F2F}
\definecolor{bsmblue}{HTML}{1976D2}
\usepackage{slashed}
\usepackage[left=1.6cm,right=1.6cm,top=2.cm,bottom=2.25cm]{geometry}

\begin{document}

\title{Trilepton signatures of heavy neutrinos in left-right symmetric models with inverse seesaw}

\author{Tanumoy Mandal}
\email{tanumoy@iisertvm.ac.in}
\affiliation{Indian Institute of Science Education and Research Thiruvananthapuram, Vithura, Kerala, 695 551, India}

\author{Subhadip Mitra}
\email{subhadip.mitra@iiit.ac.in}
\affiliation{Center for Computational Natural Sciences and Bioinformatics, International Institute of Information Technology, Hyderabad 500 032, India}
\affiliation{Center for Quantum Science and Technology, International Institute of Information Technology, Hyderabad 500 032, India}

\author{Naveen Reule}
\email{naveenreule20@iisertvm.ac.in}
\affiliation{Indian Institute of Science Education and Research Thiruvananthapuram, Vithura, Kerala, 695 551, India}

\author{Jacob Cherry Sam}
\email{jacob.cherrysam@email.ucr.edu}
\affiliation{Department of Physics and Astronomy, University of California, Riverside, California 92521, USA}

\author{Rachit Sharma}
\email{rachit21@iisertvm.ac.in}
\affiliation{Indian Institute of Science Education and Research Thiruvananthapuram, Vithura, Kerala, 695 551, India}
\affiliation{School of Physics and Institute for Collider Particle Physics, University of the Witwatersrand, Wits, 2050, Johannesburg, South Africa}
\affiliation{SGTB Khalsa College, University of Delhi, Delhi, India-110007}

\begin{abstract}
Collider searches for left-right symmetry usually look for the Keung-Senjanovi\'c (KS) process, where a heavy right-handed neutrino $N_R$ decays through a virtual $W_R$ boson into a lepton and jets. In a left-right symmetric model with an inverse seesaw, $N_R$ is pseudo-Dirac and decays mainly through its mixing with light neutrinos. The cascade $W_R \to e N_R \to e e W$ with a leptonic $W$ decay produces $eee$ and $ee\mu$ final states with missing transverse momentum, a trilepton signature that the KS process cannot generate. Using the $W$ mass constraint on the missing neutrino, we reconstruct both $M_{N_R}$ and $M_{W_R}$. At the High-Luminosity LHC, we show that suitable cuts on these reconstructed masses can reduce the Standard Model background significantly. We map the $2\sigma$ exclusion and $5\sigma$ discovery reach in the $(M_{W_R}, M_{N_R})$ plane as a function of the $SU(2)_R$ coupling $g_R$. Despite the smaller leptonic $W$ branching fraction, combining the two channels for $g_R = g_L$ excludes $W_R$ masses up to about $6.2$~TeV and reaches about $6.0$~TeV for discovery, beyond Run~2 dilepton-dijet limits reinterpreted for this model.
\end{abstract}

\maketitle

\section{Introduction}
\label{sec:intro}
\noindent
The Standard Model (SM) treats left- and right-handed fermions differently and has no right-handed neutrinos, so it cannot account for the neutrino masses required by oscillation data. The left-right symmetric model (LRSM)~\cite{Pati:1974yy,Mohapatra:1974hk,Mohapatra:1974gc,Senjanovic:1975rk,Senjanovic:1978ev} addresses both by extending the electroweak gauge group to $SU(2)_L\otimes SU(2)_R\otimes U(1)_{B-L}$, so that parity is a symmetry of the Lagrangian until $SU(2)_R$ breaks spontaneously at the scale $v_R$. The gauge structure requires right-handed neutrinos $N_R$ as partners of the right-handed charged leptons. The same symmetry breaking that gives mass to the new gauge bosons $W_R^\pm$ and $Z_R$ also gives mass to these heavy neutrinos and, through a seesaw mechanism, to the light ones~\cite{Mohapatra:1979ia,Mohapatra:1980yp}. If $v_R$ is in the TeV range, the LHC can produce these states.

The best-studied collider signature of the model was identified by Keung and Senjanovi\'c (KS)~\cite{Keung:1983uu}. A $W_R$ boson produced through the Drell-Yan process decays to a charged lepton and a heavy neutrino, which then decays to a second lepton and two jets via a virtual $W_R$. If $N_R$ is a Majorana particle, the two leptons carry the same electric charge in half of the events. This final state has almost no SM background~\cite{Ferrari:2000sp,Maiezza:2010ic,Tello:2010am}. Several later studies estimated the reach of this signal~\cite{Han:2012vk,Mitra:2016kov,Nemevsek:2018bbt}; searches for heavy neutrinos at colliders are reviewed in Refs.~\cite{Atre:2009rg,Deppisch:2015qwa,Cai:2017mow}. ATLAS and CMS have searched for this signal in the full Run~2 data set. For equal $SU(2)$ couplings ($g_R=g_L$), CMS excludes $W_R$ masses up to $4.7$~TeV in the electron channel and $5.0$~TeV in the muon channel at $M_{N_R}=M_{W_R}/2$~\cite{CMS:2021dzb}; ATLAS excludes masses up to about $6.4$~TeV for sub-TeV heavy neutrinos (Majorana or Dirac)~\cite{ATLAS:2023cjo}.

In the minimal LRSM, the heavy neutrino masses are generated by the vacuum expectation value of an $SU(2)_R$ triplet, and the neutrinos are Majorana particles. If the scalar sector contains only doublets, there is no renormalisable Majorana mass term for $N_R$, and the light neutrino masses must come from a different mechanism. A well-motivated choice is the inverse seesaw (ISS)~\cite{Mohapatra:1986aw,Mohapatra:1986bd,GonzalezGarcia:1988rw,Dev:2009aw}, which contains additional gauge-singlet fermions $S_L$; the light neutrino masses are proportional to a small lepton-number-violating parameter $\mu_S$. Together, $N_R$ and $S_L$ form heavy pseudo-Dirac pairs of Majorana states whose masses differ by $\mathcal{O}(\mu_S)$, so lepton number is conserved to a good approximation and the same-sign dilepton signal is suppressed~\cite{Kersten:2007vk,Anamiati:2016uxp,Das:2017hmg}. The heavy neutrino behaves as a Dirac particle, and the charges of the final-state leptons are fixed by the charge of the parent $W_R$. The mixing between heavy and light neutrinos can also be sizeable in this setup, in which case $N_R$ decays mostly through $N_R\to\ell W$, $\nu Z$, and $\nu h$ instead of through the virtual $W_R$~\cite{ThomasArun:2021rwf,Banerjee:2015hnl, Chen:2013foz}. The search strategy has to change accordingly, and those limits do not apply directly. They assume that $N_R$ decays to $\ell jj$ with a branching ratio close to one, whereas here the $\ell\ell jj$ final state arises only from $N_R\to\ell W$ followed by $W\to jj$, with a combined branching ratio of about one third. Reinterpreted for this model, the CMS and ATLAS searches exclude $W_R$ masses up to about $4.1$ and $5$~TeV, respectively, for $M_{N_R}=M_{W_R}/2$.

In Ref.~\cite{ThomasArun:2021rwf}, we studied this scenario for a hadronic decay of the $W$ boson produced by $N_R$, which is reconstructed as a boosted jet, and found that a $W_R$ with a mass of up to about $6$~TeV could be discovered at the High-Luminosity LHC (HL-LHC). Here we turn to the complementary case in which the $W$ boson decays to an electron or a muon. The final state then contains three charged leptons and missing transverse momentum, which cannot arise from the KS process, so a trilepton signal would by itself show that the heavy neutrino decays through its mixing with the light neutrinos. Although its rate is generally lower than that of the hadronic mode, the signature does not rely on jet substructure and avoids the large QCD background that affects hadronic final states at high $p_T$. Trilepton signals of heavy neutrinos have been discussed before~\cite{Mitra:2024eft}, both in the LRSM~\cite{Chen:2011de,Chen:2013foz} and in ISS models~\cite{Das:2012ze, Helo:2018lrsm}. CMS has also searched for heavy neutral leptons produced through the SM $W$ boson in three-lepton final states, whether Majorana or Dirac~\cite{CMS:2018iaf,CMS:2024xdq}. 

Here we take a different approach. Because lepton number is essentially conserved, the cascade $W_R\to\ell N_R\to\ell\ell W\to\ell\ell\ell\nu$ can be reconstructed almost completely. The charges of the leptons identify the one that comes from the $N_R$ decay, and the $W$ mass constraint fixes the longitudinal momentum of the neutrino up to a twofold ambiguity once the missing transverse momentum is attributed to it. The masses of $N_R$ and $W_R$ can then be computed event by event. For a signal, the reconstructed $N_R$ mass shows a narrow peak at the true value, whereas the SM backgrounds contain a genuine $W$ boson but no heavy neutrino, and populate mostly lower masses.

We use the reconstructed $N_R$ and $W_R$ masses, together with standard kinematic variables, in a rectangular selection and estimate the sensitivity of the HL-LHC~\cite{CidVidal:2018eel} for $\sqrt{s}=14$~TeV and an integrated luminosity of $3~\mathrm{ab}^{-1}$. We consider a selection with fixed thresholds, which covers the whole mass plane, and a mass-dependent selection in which the thresholds on the reconstructed masses follow the assumed masses. We keep the gauge coupling $g_R$ as a free parameter, since the production rate is proportional to $g_R^2$ and $g_R\neq g_L$ is possible if parity is broken at a higher scale~\cite{Chang:1984prl,Chang:1984prd,Borah:2010dpb,Deppisch:2015lsl}. The results are given as maps of the smallest coupling that can be excluded at $2\sigma$ or discovered at $5\sigma$ in the plane of $M_{W_R}$ and $M_{N_R}$, for the $eee$ and $ee\mu$ channels and for their combination. For $g_R=g_L$, the combined reach of the mass-dependent selection in $M_{W_R}$ is about $6.2$~TeV for exclusion and $6.0$~TeV for discovery, well beyond the Run~2 $\ell\ell jj$ limits reinterpreted for this model. We regard the trilepton channel as a complementary probe. Its backgrounds and systematic uncertainties are different. Should a signal appear in the $\ell\ell jj$ channel, this reconstruction would give an independent measurement of the heavy neutrino and $W_R$ masses in a final state that is specific to mixing-induced decays of the heavy neutrino.

The paper is organised as follows. In Sec.~\ref{sec:model} we describe the model and the properties of the heavy neutrinos that matter for the signal. Section~\ref{sec:signal} defines the trilepton signal and our benchmark points, and Sec.~\ref{sec:reconstruction} presents the reconstruction of the neutrino momentum and of the $N_R$ and $W_R$ masses together with its performance. The simulation setup is summarised in Sec.~\ref{sec:simulation}. Section~\ref{sec:selection} describes the SM backgrounds and the event selection, and Sec.~\ref{sec:results} presents the expected significances and the reach maps. We discuss the limitations of the study and possible improvements in Sec.~\ref{sec:discussion} and conclude in Sec.~\ref{sec:conclusions}.

\section{The left-right symmetric model with an inverse seesaw}
\label{sec:model}

\subsection{Gauge structure and symmetry breaking}
\noindent
We work in an LRSM with scalar doublets and an ISS sector (see Table~\ref{tab:lrsm_content}). Left- and right-handed fermions transform as doublets of $SU(2)_L$ and $SU(2)_R$, respectively, and the right-handed lepton doublet $L_R=(N_R,e_R)^T$ contains the heavy neutrino. The scalar sector consists of the bidoublet $\Phi$ and the two doublets $\chi_L$ and $\chi_R$ with unit $U(1)_{B-L}$ charges. In the fermion sector, we add three gauge singlets $S_L$ to realise the ISS mechanism.

\begin{table}[!t]
\caption{\label{tab:lrsm_content}Field content of the LRSM with scalar doublets and an inverse seesaw sector. The index $i=1,2,3$ labels generations, and we write $N_R\equiv\nu_R$ for the right-handed neutrinos.}
{\renewcommand{\arraystretch}{1.25}\setlength{\tabcolsep}{1.5pt}\small
\begin{tabular*}{\columnwidth}{@{\extracolsep{\fill}}lcccc}
\hline\hline
Field & $SU(3)_C$ & $SU(2)_L$ & $SU(2)_R$ & $U(1)_{B-L}$ \\
\hline
$q_L^i=\begin{pmatrix}u_L^i\\ d_L^i\end{pmatrix}$ & $\mathbf{3}$ & $\mathbf{2}$ & $\mathbf{1}$ & $1/3$ \\
$q_R^i=\begin{pmatrix}u_R^i\\ d_R^i\end{pmatrix}$ & $\mathbf{3}$ & $\mathbf{1}$ & $\mathbf{2}$ & $1/3$ \\
$L_L^i=\begin{pmatrix}\nu_L^i\\ e_L^i\end{pmatrix}$ & $\mathbf{1}$ & $\mathbf{2}$ & $\mathbf{1}$ & $-1$ \\
$L_R^i=\begin{pmatrix}N_R^i\\ e_R^i\end{pmatrix}$ & $\mathbf{1}$ & $\mathbf{1}$ & $\mathbf{2}$ & $-1$ \\
$S_L^i$ & $\mathbf{1}$ & $\mathbf{1}$ & $\mathbf{1}$ & $0$ \\
$\Phi=\begin{pmatrix}\phi_1^0&\phi_2^+\\ \phi_1^-&\phi_2^0\end{pmatrix}$ & $\mathbf{1}$ & $\mathbf{2}$ & $\mathbf{2}$ & $0$ \\
$\chi_L=\begin{pmatrix}\chi_L^+\\ \chi_L^0\end{pmatrix}$ & $\mathbf{1}$ & $\mathbf{2}$ & $\mathbf{1}$ & $1$ \\
$\chi_R=\begin{pmatrix}\chi_R^+\\ \chi_R^0\end{pmatrix}$ & $\mathbf{1}$ & $\mathbf{1}$ & $\mathbf{2}$ & $1$ \\
\hline\hline
\end{tabular*}}
\end{table}

The symmetry breaks in two steps:
\begin{equation}
\begin{gathered}
SU(2)_L \otimes SU(2)_R \otimes U(1)_{B-L} \\
\downarrow\rlap{$\,\langle\chi_R\rangle$} \\
SU(2)_L \otimes U(1)_Y \\
\downarrow\rlap{$\,\langle\Phi\rangle, \langle \chi_L \rangle$} \\
U(1)_{\rm em}\,.
\end{gathered}
\end{equation}
The doublet $\chi_R$ acquires a vacuum expectation value (VEV) $\langle\chi_R^0\rangle=v_R/\sqrt2$ in the TeV range, which breaks the left-right gauge structure down to that of the SM and gives mass to the new gauge bosons, $M_{W_R}\simeq g_R v_R/2$. Electroweak symmetry breaking happens as usual when $\Phi$ acquires a VEV $\langle\Phi\rangle$. The mixing between $W_L$ and $W_R$ is suppressed by the ratio of the electroweak and right-handed scales, so we neglect it. In principle, other heavy states (such as, a $Z'$ in anomaly-free $U(1)$ gauge extensions~\cite{Ekstedt:2016wyi,Choudhury:2020cpm,Deka:2021koh} or leptoquarks~\cite{Bhaskar:2023xkm,Duraikandan:2024kcy}, etc.) can also decay to RHNs and produce multilepton final states. However, in our case, the remaining new states, $Z_R$ and the extra scalars, do not contribute to the trilepton signal studied here. Indirect constraints from $K$ and $B$ meson mixing require $M_{W_R}\gtrsim3$~TeV in the minimal LRSM~\cite{Zhang:2007da,Bertolini:2014sua}, although the precise bound depends on the scalar spectrum and on the right-handed quark mixing. If parity is exact at the symmetry-breaking scale, $g_R=g_L\simeq0.65$. The couplings themselves are not independent: matching to hypercharge coupling implies $g_Y^{-2} = g_R^{-2} + g_{B-L}^{-2}$, that is $g_R > g_Y \approx 0.36$ with $g_{B-L}$ becoming non-perturbative as $g_R$ approaches this value.

\subsection{Inverse seesaw and the heavy neutrinos}
\noindent
Since the scalar sector contains no triplet scalars, no renormalisable interaction can generate direct Majorana masses for either $\nu_L$ or $N_R$. Therefore, the conventional type-I and type-II seesaw mechanisms of the minimal LRSM are absent here. Instead, the following interactions generate the light-neutrino masses
\begin{align}
    \mathcal{L}_\nu\supset{}&
    -\,\overline{L}_L\left(Y\,\Phi+\tilde Y\,\tilde\Phi\right)L_R-Y_S\,\overline{L}_R\,\tilde\chi_R\,S_L\nonumber\\
    &-Y_S'\,\overline{L}_L\,\tilde\chi_L\,S_L^{\,c}-\dfrac{1}{2}\,\mu_S\,\overline{S_L^{\,c}}\,S_L+\text{h.c.}\,,  
\end{align}
where $\tilde\Phi=\sigma_2\Phi^*\sigma_2$ and $\tilde\chi_{L,R}=i\sigma_2\chi_{L,R}^*$. After symmetry breaking, the scalars acquire the VEVs
\begin{equation}
\langle\Phi\rangle=\frac{1}{\sqrt2}
\begin{pmatrix}\kappa_1&0\\0&\kappa_2e^{i\alpha}\end{pmatrix},\qquad
\langle \chi_{L,R}\rangle=\frac{1}{\sqrt2}\begin{pmatrix}0\\v_{L,R}\end{pmatrix},
\end{equation}
with $\kappa_1^2+\kappa_2^2+v_L^2=v^2$, $v=246$~GeV, and $v_L\ll v\ll v_R$. Here, $\alpha$ is the relative phase between the two VEVs. In the $(\nu_L,N_R^c,S_L)$ basis, the $9\times9$ neutral lepton mass matrix is
\begin{equation}
\mathcal{M}=
\begin{pmatrix}
0&M_D^T&M_L\\
M_D&0&M_N\\
M_L^T&M_N^T&\mu_S
\end{pmatrix},
\label{eq:iss_matrix}
\end{equation}
where $M_D=(Y\kappa_1+\tilde Y\kappa_2e^{i\alpha})/\sqrt2$, $M_N=Y_S\,v_R/\sqrt2$, and $M_L=Y_S'\,v_L/\sqrt2$. Both the $\mu_S$ and $M_L$ terms violate lepton number. In the regime $\mu_S,\,M_L\ll M_D\ll M_N$, block diagonalisation of Eq.~\eqref{eq:iss_matrix} gives the light-neutrino mass matrix
\begin{align}
m_\nu \simeq&\
M_D^T\,(M_N^T)^{-1}\,\mu_S\,M_N^{-1}M_D  \nonumber\\
& -\left[M_L\,M_N^{-1}M_D+M_D^T\,(M_N^T)^{-1}M_L^T\right].
\label{eq:mnu_full}   
\end{align}
The first (inverse-seesaw) term scales as $\Theta^2\mu_S$, with $\Theta\sim M_D M_N^{-1}$ the active-heavy mixing, while the second (linear-seesaw) term scales as $\Theta M_L$, with $M_L/M_N\simeq(Y_S'/Y_S)\, v_L/v_R$. The linear-seesaw contribution is therefore controlled by the hierarchy between the VEVs of $\chi_L$ and $\chi_R$. Because the ISS term is itself suppressed by the small parameter $\mu_S$, this hierarchy must be strong. For $Y_S'\sim Y_S$ one needs $v_L/v_R\ll\mu_S/(\Theta M_N)$, i.e.\ $v_L/v_R\lesssim10^{-11}$ for $M_N\sim1$~TeV, $\Theta\sim10^{-2}$ and $\mu_S\sim\mathcal{O}(\text{keV})$. We simply assume that the product $Y's~ v_L$ is small enough, through a small $Y's$, a small $v_L$, or both, and neglect the linear-seesaw term.

We follow the Casas and Ibarra mixing parametrisation~\cite{Casas:2001sr}. We diagonalise the light mass matrix as $\hat m_\nu=V_L^T m_\nu V_L=\mathrm{diag}(m_1,m_2,m_3)$ using a unitary matrix $V_L$, which is the PMNS matrix in the basis where the charged lepton mass matrix is diagonal. Equation~\eqref{eq:mnu_full} can then be solved for the Dirac mass matrix,
\begin{equation}
M_D=M_N\,\hat\mu_S^{-1/2}\,R\,\hat m_\nu^{1/2}\,V_L^\dagger\,,
\label{eq:casas_ibarra}
\end{equation}
where $\hat\mu_S$ is the diagonal form of $\mu_S$ and $R$ is an arbitrary complex orthogonal matrix, $R^TR=1$. For any choice of $M_N$, $\hat\mu_S$, and $R$, this reproduces the measured oscillation parameters, while the three complex angles of $R$ remain unconstrained by oscillation data.

\begin{figure}[!t]
\centering
\includegraphics[width=\columnwidth]{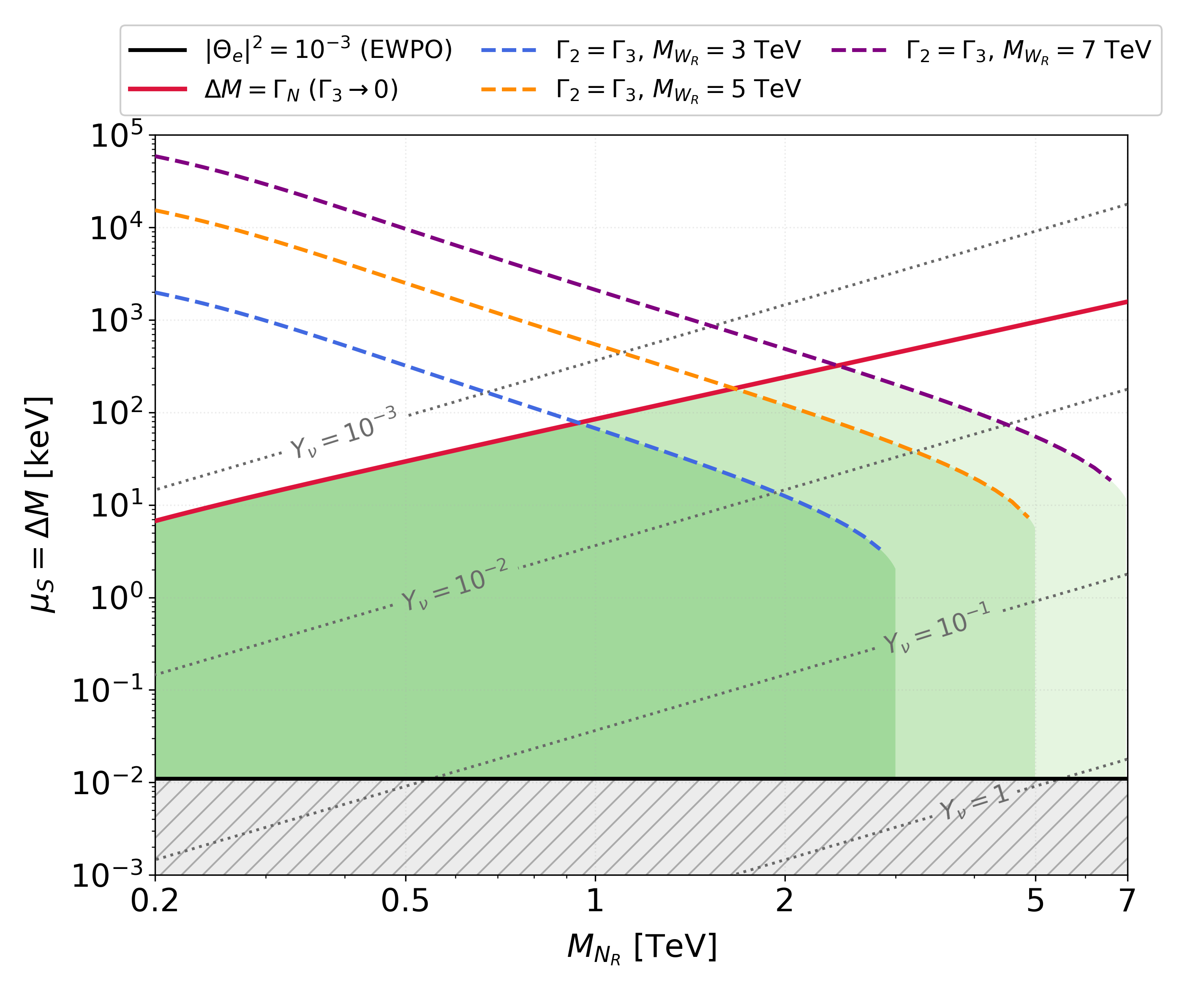}
\caption{Region of the $(M_{N_R},\mu_S)$ plane in which the heavy neutrinos are pseudo-Dirac and decay through their mixing with the light neutrinos. The first-generation heavy neutrino mixes with $\nu_e$ only, with normal ordering and $m_1=0.01$~eV. Below the solid red line, $\Delta M<\Gamma_N$ and lepton number violation is suppressed. Below the dashed lines, the two-body decays $N_R\to\ell W,\nu Z,\nu h$ dominate over the three-body decay through a virtual $W_R$, for $M_{W_R}=3$, $5$, and $7$~TeV with $g_R=g_L$; these lines end at $M_{N_R}=M_{W_R}$. The hatched region, $|\Theta_e|^2>10^{-3}$, is excluded by electroweak precision data. The shaded areas show the allowed window for each $M_{W_R}$, and the dotted lines are contours of the Dirac Yukawa coupling $Y_\nu$.}
\label{fig:mus_window}
\end{figure}

The heavy states form three pseudo-Dirac pairs, with masses and splitting
\begin{equation}
M_{N_i^\pm}\simeq M_{N_i}\pm\dfrac12\mu_{S,i}\,,\qquad \Delta M_i\simeq\mu_{S,i}\ll M_{N_i}\,,
\label{eq:pseudo_dirac}
\end{equation}
up to corrections of order $M_D^2/M_N^2$. The splitting of a pair is much smaller than the $N_R$ decay width $\Gamma_N$. A lepton-number-violating (same-sign) final state can then arise only through the interference of the two components of the pair, and its rate relative to the lepton-number-conserving one is~\cite{Anamiati:2016uxp,Godbole:2020mes}
\begin{equation}
R_{\ell\ell}\simeq\frac{\Delta M^2}{2\Gamma_N^2+\Delta M^2}\,\approx\,\frac{\Delta M^2}{2\Gamma_N^2}\qquad(\Delta M\ll\Gamma_N)\,.
\label{eq:rll}
\end{equation}
In the ISS, the light neutrino mass is of order $|\Theta|^2\mu_S$, so for fixed $m_\nu$ a smaller $\mu_S$ means a larger mixing and a larger width. The condition $\Delta M\ll\Gamma_N$ is therefore met comfortably for $\mu_S$ at the keV scale, $R_{\ell\ell}$ is negligible, and the heavy neutrino behaves as a Dirac particle. The signal studied in this work is the lepton-number-conserving trilepton channel.

The decays $N_R\to\ell W$, $\nu Z$, and $\nu h$ proceed through the active-heavy mixing, which follows from Eq.~\eqref{eq:casas_ibarra},
\begin{equation}
|\Theta_{\alpha i}|^2\simeq\big|(M_N^{-1}M_D)_{i\alpha}\big|^2=\Big|\big(\hat\mu_S^{-1/2}\,R\,\hat m_\nu^{1/2}\,V_L^\dagger\big)_{i\alpha}\Big|^2\,.
\label{eq:mixing}
\end{equation}
For $M_{N_R}\gg M_W$, the branching ratios of these modes follow the pattern expected from Goldstone boson equivalence, $\mathrm{BR}(\ell W):\mathrm{BR}(\nu Z):\mathrm{BR}(\nu h)\simeq2:1:1$, independently of the size of the mixing. The full phase-space expressions give $\mathrm{BR}(N_R\to\ell W)$ between $0.5$ and $0.6$ for $M_{N_R}\geq200$~GeV, so using the asymptotic value of $0.5$ is slightly conservative.

This pattern holds as long as the mixing-induced modes dominate over the three-body decay $N_R\to\ell q\bar q'$ through a virtual $W_R$, whose width does not depend on $\Theta$. Figure~\ref{fig:mus_window} shows the region of the $(M_{N_R},\mu_S)$ plane in which this is the case and the heavy neutrinos behave as Dirac particles. We take the first-generation heavy neutrino to mix with $\nu_e$ only, which fixes the first row of $R$ up to normalisation, and use normal ordering with $m_1=0.01$~eV, so that $|\Theta_e|^2\propto1/\mu_S$. Three conditions define the region. First, lepton number violation must be negligible, $\Delta M\ll\Gamma_N$. Since $\Gamma_N\propto|\Theta|^2M_N^3\propto M_N^3/\mu_S$ while $\Delta M\simeq\mu_S$, this bounds $\mu_S$ from above. Second, the two-body decays must dominate over the three-body decay. The ratio of their widths scales as $\mu_SM_N^2/M_{W_R}^4$, so this condition also bounds $\mu_S$ from above. Third, the mixing must satisfy the constraints from electroweak precision data and lepton flavour universality, $|\Theta_e|^2\lesssim10^{-3}$~\cite{Fernandez-Martinez:2016lgt}, which bounds $\mu_S$ from below at about $10$~eV. For all masses considered in this work, the three conditions are met for $\mu_S$ between about $10$~eV and a few keV. In this window, the Dirac Yukawa coupling $Y_\nu=\sqrt2M_D/v$ ranges from about $10^{-3}$ to $0.5$, many orders of magnitude above the value $\sqrt{2m_\nu M_N}/v\sim10^{-6}$ that a type-I seesaw would need for TeV-scale heavy neutrinos.

\section{Trilepton signal and benchmark points}
\label{sec:signal}
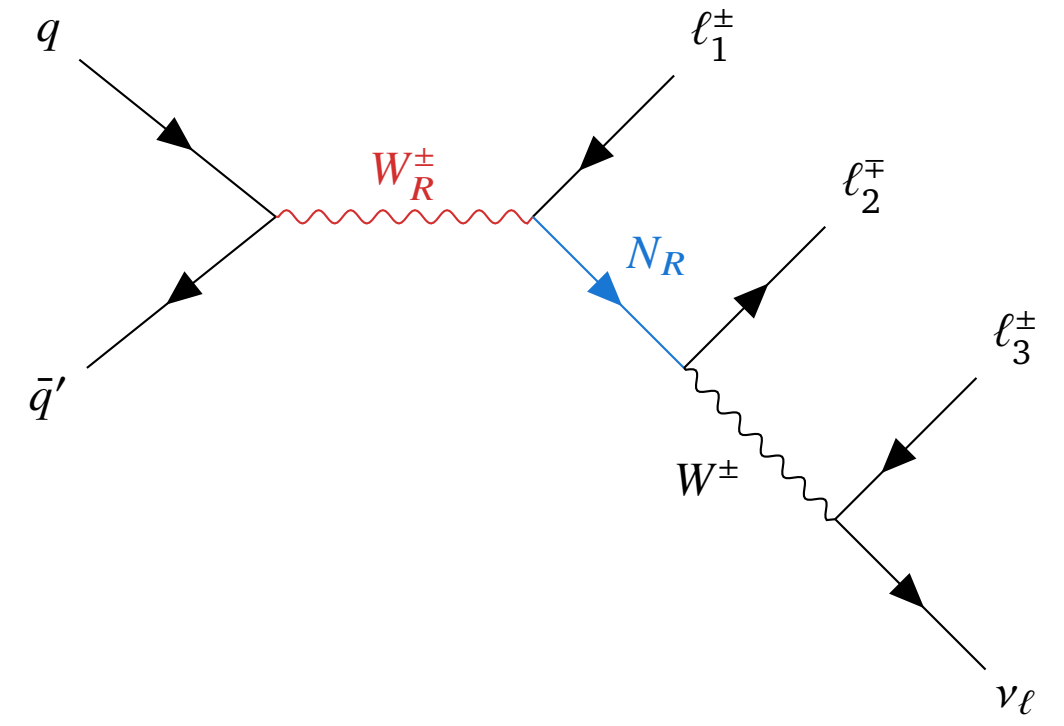
\begin{figure}[!t]
\centering
\begin{tikzpicture}
\begin{feynman}
\vertex (i1) at (0,1.2) {$q$};
\vertex (i2) at (0,-1.2) {$\bar q'$};
\vertex (v1) at (1.5,0);
\vertex (v2) at (3.2,0);
\vertex (f1) at (4.4,1.2) {$\ell_1^\pm$};
\vertex (v3) at (4.2,-1.0);
\vertex (f2) at (5.4,0.2) {$\ell_2^\mp$};
\vertex (v4) at (5.2,-2.0);
\vertex (f3) at (6.4,-0.8) {$\ell_3^\pm$};
\vertex (f4) at (6.4,-3.2) {$\nu_\ell$};
\diagram* {
(i1) -- [fermion] (v1) -- [fermion] (i2),
(v1) -- [boson, bsmred, edge label={\color{bsmred}$W_R^\pm$}] (v2),
(v2) -- [anti fermion] (f1),
(v2) -- [fermion, bsmblue, edge label={\color{bsmblue}$N_R$}] (v3),
(v3) -- [fermion] (f2),
(v3) -- [boson, edge label'={$W^\pm$}] (v4),
(v4) -- [anti fermion] (f3),
(v4) -- [fermion] (f4),
};
\end{feynman}
\end{tikzpicture}
\caption{Trilepton signal in the LRSM with a pseudo-Dirac heavy neutrino. The charges of the leptons satisfy $q(\ell_1)=q(\ell_3)=-q(\ell_2)$.}
\label{fig:trilepton_feynman}
\end{figure}

\noindent
We probe the model through the Drell-Yan production of a heavy charged gauge boson at the $14$~TeV HL-LHC, followed by the decay chain
\begin{equation}
pp\to W_R^\pm\to\ell_1^\pm N_R\to\ell_1^\pm\ell_2^\mp W^\pm\to\ell_1^\pm\ell_2^\mp\ell_3^\pm\nu\,,
\label{eq:process}
\end{equation}
as illustrated in Fig.~\ref{fig:trilepton_feynman}. We consider the mass hierarchy $M_W<M_{N_R}<M_{W_R}$, which allows the successive two-body decays of the on-shell $W_R$, $N_R$, and $W$. The $W_R$ decays into a charged lepton and a heavy neutrino, which subsequently decays into a second charged lepton and a SM $W$ boson. The latter undergoes a leptonic decay, resulting in three charged leptons and a neutrino in the final state. Since lepton number is approximately conserved in the pseudo-Dirac limit, the charged lepton from the $N_R$ decay has the opposite charge to that of the parent $W_R$, whereas the leptons originating from the $W_R$ and $W$ decays carry the same charge. The characteristic signature is therefore $\ell^\pm\ell^\mp\ell^\pm$ accompanied by missing transverse momentum from the undetected neutrino.

Requiring three isolated charged leptons with high transverse momenta suppresses most SM backgrounds, making the trilepton final state promising. The undetected neutrino gives rise to missing transverse momentum, from which we infer its transverse momentum. Imposing the on-shell $W$-boson mass constraint then determines its longitudinal momentum up to a twofold ambiguity, allowing us to reconstruct the invariant masses of both $N_R$ and $W_R$.

The hadronic decay mode of the $W$ boson, studied in Ref.~\cite{ThomasArun:2021rwf}, has a branching fraction approximately three times larger than the combined $W\to e\nu$ and $W\to\mu\nu$ modes. However, when the $W$ boson is highly boosted, its hadronic decay products can merge into a single large-radius jet, requiring dedicated jet-substructure techniques to distinguish it from ordinary QCD jets~\cite{Dey:2022fjt,Arun:2022zpn,Mandal:2023dlz}. The leptonic channel avoids this complication and provides a complementary search strategy, although at the cost of a smaller branching fraction.

For our analysis, we assume that only the first-generation right-handed neutrino (RHN) is lighter than $W_R$ and can therefore be produced in its on-shell decay. Consequently, the charged leptons originating from the $W_R$ and $N_R$ decays are both electrons, $\ell_1=\ell_2=e$. The $W$ boson subsequently decays into either an electron or a muon, defining the two trilepton channels, $eee$ and $ee\mu$, considered in this study. We do not include $W\to\tau\nu$ decays.

The expected signal yield is given by
\begin{align}
N_S ={} &\sigma(pp\to W_R)\times\mathrm{BR}(W_R\to eN_R) \nonumber\\
&\times\mathrm{BR}(N_R\to eW)\times\mathrm{BR}(W\to\ell\nu)\nonumber\\
&\times\mathrm{CE}_S\times \mathcal L\,,
\label{eq:signal_yield}
\end{align}
where $\mathcal L$ is the integrated luminosity and $\mathrm{CE}_S$ denotes the total signal selection efficiency, including detector acceptance. The $W_R$ production cross section is obtained from the leading-order calculation described in Sec.~\ref{sec:simulation}, supplemented by a constant QCD $K$ factor of $1.3$. This choice is consistent with the NLO QCD corrections, which typically yield $K$ factors between $1.2$ and $1.4$ for Drell-Yan $W_R$ production in the multi-TeV mass range~\cite{Mitra:2016kov,Mattelaer:2016ynf}. Higher-order threshold-resummation effects can further enhance the cross section at larger $W_R$ masses~\cite{Mitra:2016kov}.

The branching fraction $\mathrm{BR}(W_R\to eN_R)$ is evaluated at the tree level, assuming that only the first-generation RHN is kinematically accessible in decays of $W_R$. For the subsequent decay, we adopt $\mathrm{BR}(N_R\to eW)=0.5$, corresponding to the asymptotic branching-ratio pattern $\mathrm{BR}(N_R\to eW):\mathrm{BR}(N_R\to\nu Z):\mathrm{BR}(N_R\to\nu h)\simeq 2:1:1$. We use $\mathrm{BR}(W\to e\nu)=\mathrm{BR}(W\to\mu\nu)=10.86\%$~\cite{ParticleDataGroup:2024cfk}.

In the narrow-width approximation, and assuming that the mixing-induced two-body decays of $N_R$ dominate, the branching fractions and selection efficiencies are approximately independent of $g_R$. Consequently, the signal yield scales as $N_S\propto g_R^2$, allowing us to extrapolate the results obtained at a reference coupling to other values of $g_R$, as described in Sec.~\ref{sec:results}. The validity and limitations of the narrow-width approximation are discussed in Sec.~\ref{sec:discussion}.

Two benchmark points are used throughout the paper, both with the SM-like coupling $g_R=0.65$. Benchmark~A has $(M_{W_R},M_{N_R})=(3~\mathrm{TeV},1~\mathrm{TeV})$. It lies well inside the reach of the HL-LHC and gives a large event sample, with an effective cross section $\sigma\times\mathrm{BR}=0.79$~fb in each channel. Benchmark~B has $(5~\mathrm{TeV},2.5~\mathrm{TeV})$, with $\sigma\times\mathrm{BR}=0.012$~fb. It lies close to the present $\ell\ell jj$ limit reinterpreted for our model and illustrates the multi-TeV region in which the HL-LHC sensitivity is limited by the number of signal events. In the rest frame of the $W_R$, the primary lepton carries a momentum $(M_{W_R}^2-M_{N_R}^2)/(2M_{W_R})$, which is $1.33$~TeV for Benchmark~A and $1.88$~TeV for Benchmark~B. This hard lepton is the first characteristic feature of the signal, as can be seen in the leading lepton $p_T$ distributions of Figs.~\ref{fig:dist_eee} and~\ref{fig:dist_eemu}.

\section{Event reconstruction}
\label{sec:reconstruction}
\noindent
The undetected neutrino in the final state prevents a direct reconstruction of the $N_R$ and $W_R$ invariant masses using only the visible decay products. However, in the lepton-number-conserving channel, the charge correlations among the three leptons, together with the on-shell $W$-boson mass constraint, allow the decay chain in Eq.~\eqref{eq:process} to be reconstructed event by event, up to a limited kinematic ambiguity. We denote the charged leptons originating from the decays $W_R\to\ell_{W_R}N_R$, $N_R\to\ell_N W$, and $W\to\ell_W\nu$ by $\ell_{W_R}$, $\ell_N$, and $\ell_W$, respectively, corresponding to $\ell_1$, $\ell_2$, and $\ell_3$ in Eq.~\eqref{eq:process}.

\subsection{Lepton assignment}
\label{subsec:assignment}
\noindent
For the lepton-number-conserving decay chain, charge conservation implies $q(\ell_{W_R})=q(\ell_W)=-q(\ell_N)$. Thus, $\ell_N$ can be uniquely identified as the lepton whose charge is opposite to that of the other two. In the $ee\mu$ channel, the muon must originate from the $W$ decay, while the two electrons can be distinguished by their charges: the electron with charge opposite to that of the muon is $\ell_N$, and the other is $\ell_{W_R}$. The lepton assignment is therefore unambiguous. In the $eee$ channel, although $\ell_N$ is uniquely identified, the two same-sign electrons can originate from either the $W_R$ or the $W$ decay, resulting in a twofold combinatorial ambiguity.

For each lepton assignment, the on-shell $W$-boson mass constraint yields a quadratic equation for the longitudinal momentum of the neutrino. Consequently, before resolving the ambiguities, there are up to four kinematic solutions in the $eee$ channel and two in the $ee\mu$ channel. Rather than relying on the transverse-momentum ordering of the leptons, we resolve the combinatorial ambiguity in the $eee$ channel using the transverse mass $M_T(\ell,\slashed{E}_T)$. Of the two same-sign electrons, the one with the smaller transverse mass is assigned to $\ell_W$, as motivated by the kinematic condition in Eq.~\eqref{eq:mt_condition}.

\subsection{Neutrino momentum}
\label{subsec:neutrino}
\noindent
Assuming that the missing transverse momentum arises entirely from the final-state neutrino, we set $\vec p_T^{\,\nu}=\vec p_T^{\,\rm miss}$. The longitudinal component of the neutrino momentum can then be determined, up to a twofold ambiguity, by imposing the on-shell $W$-boson mass constraint,
\begin{equation}
(p_{\ell_W}+p_\nu)^2=M_W^2\,.
\label{eq:w_mass_constraint}
\end{equation}
Neglecting the charged-lepton mass, this leads to a quadratic equation for $p_z^\nu$, with solutions (see Appendix~\ref{app:neutrino})
\begin{equation}
p_z^{\nu,\pm}=\frac{a\,p_z^{\ell_W}\pm E_{\ell_W}\sqrt{a^2-4\,p_T^2(\ell_W)\,(\slashed{E}_T)^2}}{2\,p_T^2(\ell_W)}\,,
\label{eq:pz_roots}
\end{equation}
where $a=M_W^2+2\,\vec p_T^{\,\ell_W}\!\cdot\vec p_T^{\,\rm miss}$. The solutions are real when the discriminant is nonnegative, which is equivalent to requiring
\begin{align}
M_T(\ell_W,\slashed{E}_T)&=\sqrt{2\,p_T(\ell_W)\,\slashed{E}_T\,(1-\cos\Delta\phi_{\ell_W,\nu})} \le M_W.
\label{eq:mt_condition}
\end{align}

This condition also provides a useful criterion for resolving the lepton-assignment ambiguity in the $eee$ channel. For the electron originating from the $W$ decay, the transverse mass is kinematically bounded by $M_W$ at parton level, producing the characteristic Jacobian edge in its distribution. Detector resolution and the finite $W$ width can smear this endpoint. In contrast, the electron from the primary $W_R$ decay typically carries a much larger transverse momentum and is not constrained by the $W$-boson mass. Its transverse mass with the missing momentum can therefore exceed $M_W$, leading to complex solutions for $p_z^\nu$. We consequently assign the same-sign electron with the smaller transverse mass to $\ell_W$.

In the $ee\mu$ channel there is no ambiguity of this kind, but the condition is not always satisfied by the true lepton pair. For some benchmark points the roots can become complex in some of the signal events, because resolution effects or the finite $W$ width push the true pair above $M_W$. In that case, we take the real part of the solution,
\begin{equation}
p_z^\nu=\frac{a\,p_z^{\ell_W}}{2\,p_T^2(\ell_W)}\,,
\label{eq:pz_real}
\end{equation}
which corresponds approximately to $\eta_\nu=\eta_{\ell_W}$. When both roots are real, we use the one with the smaller $|p_z^\nu|$, which avoids the large rapidity tail of the other.

\subsection{Reconstructed masses and performance}
\label{subsec:performance}
\noindent
With the lepton assignment and the neutrino momentum fixed, we define
\begin{align}
M^{\mathrm{reco}}_{N_R}&=\sqrt{(p_{\ell_N}+p_{\ell_W}+p_\nu)^2}\,,\nonumber\\
M^{\mathrm{reco}}_{W_R}&=\sqrt{(p_{\ell_{W_R}}+p_{\ell_N}+p_{\ell_W}+p_\nu)^2}\,,
\label{eq:masses}
\end{align}
the invariant masses of the decay products of $N_R$ and $W_R$. Both are used in the selection of events in Sec.~\ref{sec:selection}. The $W_R$ mass is the invariant mass of all three leptons and the neutrino. It depends on the lepton assignment only through the neutrino solution and is therefore the more robust of the two.

We tested the reconstruction on generator-matched signal events at multiple benchmarks, and the lepton assignment is almost always correct, and in the $ee\mu$ channel the flavour of the muon removes the combinatorial ambiguity entirely. The lower $|p_z^\nu|$ root is the true one in only slightly more than half of the events, but the two roots lead to similar masses, so the twofold ambiguity costs little. With the lower root, the $N_R$ mass is reconstructed with a median error of about $3\%$ in both channels. For multi-TeV $W_R$ bosons, the reconstructed $W_R$ mass is sharply peaked, with a median within about $1\%$ of the true mass after the selection of Sec.~\ref{subsec:selection}. The resulting distributions of $M^{\mathrm{reco}}_{N_R}$ and $M^{\mathrm{reco}}_{W_R}$ are shown in Figs.~\ref{fig:dist_eee} and~\ref{fig:dist_eemu}, together with other kinematic variables.

\section{Simulation setup}
\label{sec:simulation}
\noindent
We implement the interactions of the LRSM with the ISS sector in \textsc{FeynRules}~\cite{Alloul:2013bka} and export the model in the UFO format~\cite{Degrande:2011ua}. Signal and background events are generated at leading order with \textsc{MadGraph5\_aMC@NLO}~\cite{Alwall:2014hca}, using the \textsc{NNPDF2.3} leading-order parton distribution functions~\cite{Ball:2012cx} and the default dynamic scale choice. The parton-level events are passed to \textsc{Pythia8}~\cite{Sjostrand:2014zea} for parton showering and hadronisation, and then to \textsc{Delphes3}~\cite{deFavereau:2013fsa} for the detector simulation, using the standard CMS detector card. Jets are clustered with the anti-$k_T$ algorithm~\cite{Cacciari:2008gp} as implemented in \textsc{FastJet}~\cite{Cacciari:2011ma}. The signal does not contain jets at leading order, so apart from the $b$-jet veto of Sec.~\ref{sec:selection}, we impose no requirement on jets. No pileup is simulated.

For the signal, we generate a grid with $M_{W_R}$ between $3$ and $8$~TeV and $M_{N_R}$ between $0.1$~TeV and $M_{W_R}$. The reach maps of Sec.~\ref{sec:results} are interpolated on this grid. The signal is normalised as described in Sec.~\ref{sec:signal}. The background samples are generated at leading order, with up to 2 additional partons for $WZ$, $t\bar tW^\pm$, $t\bar t$, and $Z$+jets. Each sample is normalised to its leading-order cross section, including the generator-level requirements, multiplied by a $K$ factor equal to the ratio of the inclusive higher-order and leading-order cross sections at $14$~TeV. The $K$ factors are listed in Table~\ref{tab:kfactors}.

\begin{table}[!t]
\caption{\label{tab:kfactors}Inclusive leading-order cross sections at $\sqrt{s}=14$~TeV from \textsc{MadGraph5\_aMC@NLO}, higher-order (HO) cross sections from the literature, and the resulting $K$ factors applied to the background samples. The HO values are for on-shell $WZ$, $ZZ$, and triple gauge bosons, with NLO QCD and NLO EW corrections for $WWW$ and $WWZ$, and with NLO EW corrections included in the a(pproximate-)N$^3$LO top quark results.}
{\renewcommand{\arraystretch}{1.25}\setlength{\tabcolsep}{1.5pt}\small
\begin{tabular*}{\columnwidth}{@{\extracolsep{\fill}}lrrrl}
\hline\hline
Process & $\sigma_{\rm LO}$ (pb) & $\sigma_{\rm HO}$ (pb) & $K$ & Ref. \\
\hline
$W^\pm Z$ & 34.2 & 56.85 (NNLO) & 1.66 & \cite{Grazzini:2016swo} \\
$ZZ$ & 10.26 & 18.20 (NNLO) & 1.77 & \cite{Azzi:2019yne} \\
$t\bar t$ & 576 & 987 (aN$^3$LO) & 1.72 & \cite{Kidonakis:2023juy} \\
$t\bar tW^\pm$ & 0.506 & 0.905 (aN$^3$LO) & 1.78 & \cite{Kidonakis:2023jpj} \\
$t\bar tZ$ & 0.705 & 1.211 (aN$^3$LO) & 1.72 & \cite{Kidonakis:2024lht} \\
$WWW$ & 0.125 & 0.242 (NLO) & 1.93 & \cite{Dittmaier:2017bnh} \\
$WWZ$ & 0.094 & 0.203 (NLO) & 2.16 & \cite{Nhung:2013jta} \\
$Z/\gamma^*\to e^+e^-$ & 940.34 & 987.35 (NNLO) & 1.05 & \cite{Armadillo:2024ncf} \\
\hline\hline
\end{tabular*}}
\end{table}

\section{Backgrounds and event selection}
\label{sec:selection}
\noindent
We study the signals $pp\to W_R^\pm\to e^\pm N_R\to e^\pm e^\mp e^\pm\nu$ ($eee$) and $pp\to W_R^\pm\to e^\pm N_R\to e^\pm e^\mp\mu^\pm\nu$ ($ee\mu$) at the HL-LHC, with $\sqrt{s}=14$~TeV and $\mathcal{L}=3000~\mathrm{fb}^{-1}$.

\subsection{Standard Model backgrounds}
\label{subsec:backgrounds}
\noindent
We divide the SM backgrounds into irreducible and reducible ones. \emph{Irreducible backgrounds} contain three genuine prompt leptons ($eee$ or $ee\mu$) and genuine neutrinos. The dominant one is $p~p\to\ell\nu\,e^+e^-$, with $\ell=e$ for $eee$ and $\ell=\mu$ for $ee\mu$, which includes on-shell $WZ$, off-shell $Z/\gamma*$ contributions and interference and has genuine $\slashed{E}_T$ and an opposite-sign electron pair at the $Z$ pole. The $t\bar tV$ processes ($t\bar tZ$ and $t\bar tW^\pm$) have leptons and neutrinos from the decays of the top quarks and the boson, and always contain two $b$ quarks. The triple-boson processes $VVV$ ($WWW$ and $WWZ$) are suppressed by their small cross sections. Among them, $WWW\to3\ell3\nu$ has neither a $Z$ peak nor $b$ jets, which makes it the main non-resonant survivor at high mass.

\emph{Reducible backgrounds} reach the signal region through lost leptons, mismeasured $\slashed{E}_T$, or jets misidentified as leptons. The process $ZZ\to4e$ ($eee$) or $ZZ\to2e2\mu$ ($ee\mu$) enters when one lepton is lost because it falls outside the acceptance, has a transverse momentum below threshold, or fails isolation. It has no genuine neutrinos, and its $\slashed{E}_T$ comes from the lost lepton and resolution effects. Drell-Yan production with jets ($Z$+jets) has two prompt electrons at the $Z$ pole, and enters when a jet is misidentified as the third lepton (an electron in the $eee$ channel and a muon in the $ee\mu$ channel). We model this with a constant jet fake rate $f_{j\to e}=f_{j\to\mu}=10^{-4}$ as is common in phenomenological studies~\cite{Nemevsek:2016enw, Pascoli:2018heg}. Dileptonic $t\bar t$ production with jets has two prompt leptons ($e^+e^-$, or $e^\pm\mu^\mp$ in the $ee\mu$ channel), genuine $\slashed{E}_T$, and two $b$ jets, and enters through the same fake rate. Although the fake rates are small, the cross sections of $Z$+jets and $t\bar t$ production are several orders of magnitude larger than that of the signal, so these processes need to be suppressed as well.

\begin{table*}
\caption{\label{tab:cutflow_eee}Cut flow in the $eee+\slashed{E}_T$ channel. The yields are expected numbers of events for $\mathcal{L}=3000~\mathrm{fb}^{-1}$ at $\sqrt{s}=14$~TeV. The signal benchmarks A and B are $(M_{W_R},M_{N_R})=(3~\mathrm{TeV},1~\mathrm{TeV})$ and $(5~\mathrm{TeV},2.5~\mathrm{TeV})$, both with $g_R=0.65$. The $VVV$ column contains $WWW$ and $WWZ$, and $t\bar tW^\pm$ combines $t\bar tW^+$ and $t\bar tW^-$.}
{\renewcommand{\arraystretch}{1.25}\setlength{\tabcolsep}{1.5pt}\small
\begin{tabular*}{\textwidth}{@{\extracolsep{\fill}}lrrrrrrrrrr}
\hline\hline
 & \multicolumn{3}{c}{Reducible} & \multicolumn{4}{c}{Irreducible} & SM & \multicolumn{2}{c}{Signal} \\
\cmidrule(lr){2-4}\cmidrule(lr){5-8}\cmidrule(lr){10-11}
Selection & $ZZ$\,(1 lost) & $Z$+jets & $t\bar t$+jets & $WZ$ & $t\bar tZ$ & $t\bar tW^\pm$ & $VVV$ & total & A & B \\
\hline
1.~Preselection & $20480.7$ & $1277.1$ & $2442.5$ & $69552.6$ & $1900.2 $& $594.2$ & $494.8$ & $96742.2$ & $824.1 $& $12.23$ \\
2.~$p_T^{(1)}>120$~GeV & $2355.1$ & $323.3$ & $1115.0 $& $13438.2$ & $904.5 $&$ 199.9$ & $182.1$ & $18518.1$ &$ 824.1$ & $12.23$ \\
3.~$p_T^{(3)}>50$~GeV & $944.3$ & $158.4$ & $369.3$ & $5571.1$ & $319.5 $& $67.1$ & $62.5$ & $7492.2 $& $799.4$ & $12.14$ \\
4.~$\slashed{E}_T>100$~GeV & $3.6$ & $5.4$ & $116.4$ & $1481.7 $& $86.9$ & $28.4$ & $20.1$ & $1742.5$ & $719.1$ & $11.79$ \\
5.~$S_T>500$~GeV &$ 2.9$ & $3.9$ & $74.6$ & $1005.5 $& $56.4$ & $17.3$ & $14.8$ & $1175.4 $& $719.1$ & $11.79 $\\
6.~$Z$ veto & $<1$ & $3.6 $& $65.7 $& $109.8$ &$ 1.2$ & $16.2$ & $3.8 $& $201.0$ & $717.6$ & $11.78 $\\
7.~$b$-jet veto & $<1$ & $2.5$ & $11.1$ & $100.9$ & $<1$ & $3.7$ & $3.6 $& $122.8$ & $695.8$ & $11.37$ \\
8.~$M^{\mathrm{reco}}_{N_R}>200$~GeV & $<1$ & $2.5 $& $9.7$ & $86.7$ & $<1$ & $3.6$ & $3.3 $& $106.9$ & $695.8$ & $11.37$ \\
9.~$M^{\mathrm{reco}}_{W_R}>1$~TeV & $\approx 0$ & $1.1$ & $3.3$ & $21.2$ & $\approx 0$ & $< 1$ & $1.2$ & $\mathbf{27.6}$ & $\mathbf{695.5}$ & $\mathbf{11.37}$ \\
\hline\hline
\end{tabular*}}
\caption{\label{tab:cutflow_eemu}Same as Table~\ref{tab:cutflow_eee}, for the $ee\mu+\slashed{E}_T$ channel.}
{\renewcommand{\arraystretch}{1.25}\setlength{\tabcolsep}{1.5pt}\small
\begin{tabular*}{\textwidth}{@{\extracolsep{\fill}}lrrrrrrrrrr}
\hline\hline
 & \multicolumn{3}{c}{Reducible} & \multicolumn{4}{c}{Irreducible} & SM & \multicolumn{2}{c}{Signal} \\
\cmidrule(lr){2-4}\cmidrule(lr){5-8}\cmidrule(lr){10-11}
Selection & $ZZ$\,(1 lost) & $Z$+jets & $t\bar t$+jets & $WZ$ & $t\bar tZ$ & $t\bar tW^\pm$ & $VVV$ & total & A & B \\
\hline
1.~Preselection & $8381.0$ & $1277.4$ & $6086.$5 & $129448.4$ & 1$984.9$ & $1756.8$ & $992.6$ & $149927.6$ & $997.1$ & $14.77$ \\
2.~$p_T^{(1)}>120$~GeV & $680.3$ & $222.6 $& $2624.0$ & $18389.2$ & $826.3$ & $464.1 $& $255.2$ & $23461.7$ & $996.9$ & $14.77$ \\
3.~$p_T^{(3)}>50$~GeV & $194.5$ & $111.3 $& $825.0$ & $5225.1$ & $282.0$ & $164.2$ & $95.5$ & $6897.5 $& $968.3$ & $14.67 $\\
4.~$\slashed{E}_T>100$~GeV & $7.0$ & $4.7$ & $269.8$ & $1696.6$ & $76.7$ & $73.7$ & $33.8$ & $2162.4$ & $870.4$ & $14.04$ \\
5.~$S_T>500$~GeV & $4.2$ & $3.9$ & $177.8$ & $1146.1$ & $52.8$ & $47.7$ & $25.5$ & $1458.0$ & $870.4$ & $14.04$ \\
6.~$Z$ veto & $<1$ & $3.6$& $167.5$ & $157.8$ & $2.1$ & $46.5$ & $13.3$ & $391.4$ & $869.0$ & $14.03$ \\
7.~$b$-jet veto & $<1$ & $2.5$ & $29.6$ & $151.3$ & $<1$ & 12.9 & $12.6$ & $210.1$ & $844.8$ & $13.58$ \\
8.~$M^{\mathrm{reco}}_{N_R}>200$~GeV & $<1$ & $2.5$ & $14.4$ & $125.6$ & $<1$ & $7.6$ & $7.6$ & $158.9$ & $844.8$ & $13.58$ \\
9.~$M^{\mathrm{reco}}_{W_R}>1$~TeV & $\approx 0$ & $<1$ & $4.5$ & $35.4$ & $\approx 0$ & $2.2$ & $3.0$ & $\mathbf{45.7}$ & $\mathbf{844.8}$ & $\mathbf{13.58}$ \\
\hline\hline
\end{tabular*}}
\end{table*}
\begin{figure*}
\centering
\includegraphics[width=\textwidth]{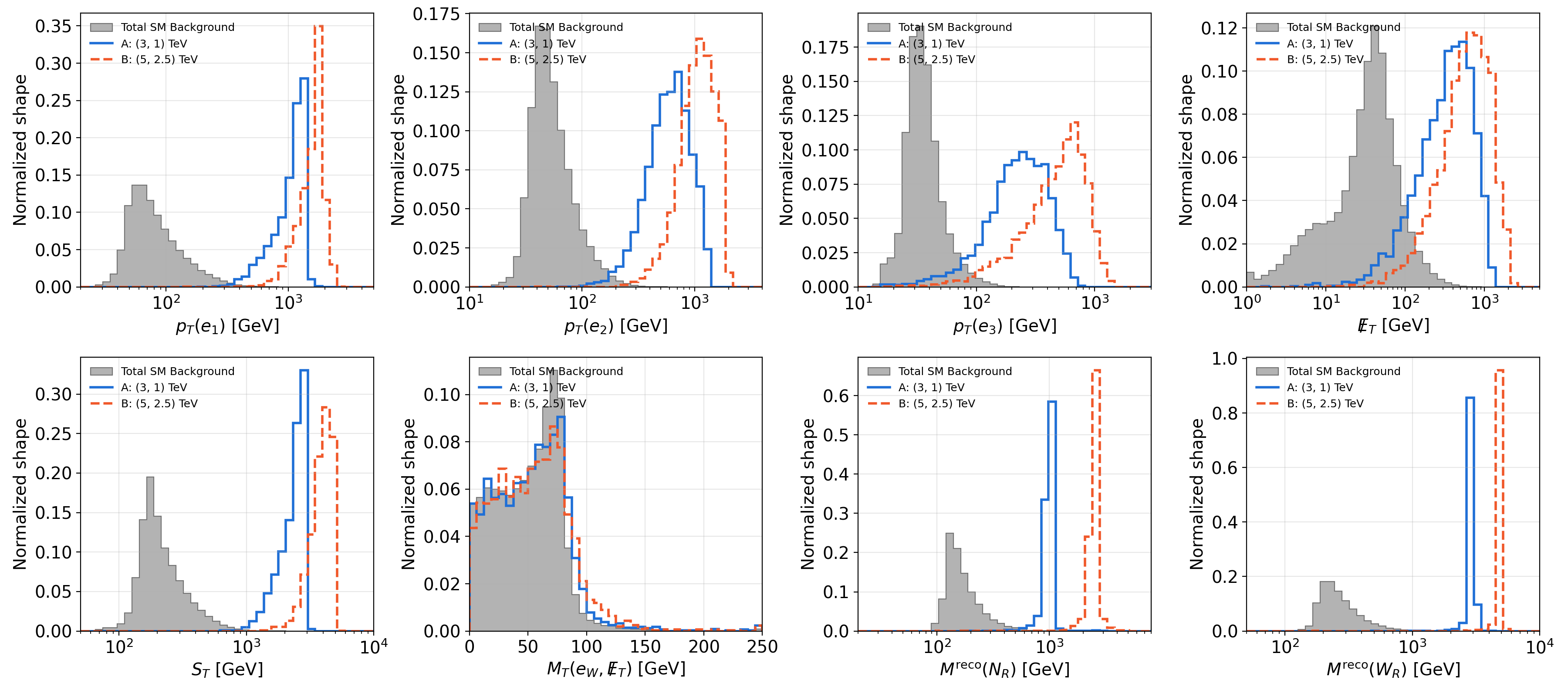}
\caption{\label{fig:dist_eee} Normalised shapes of kinematic variables in the $eee$ channel after the preselection, for Benchmark~A (solid blue), Benchmark~B (dashed orange), and the total SM background (grey histogram). From left to right and from top to bottom: the transverse momenta of the leading, subleading, and third electron, $\slashed{E}_T$, $S_T$, $M_T(e_W, \slashed{E}_T)$, and the reconstructed $N_R$ and $W_R$ masses.}

\includegraphics[width=\textwidth]{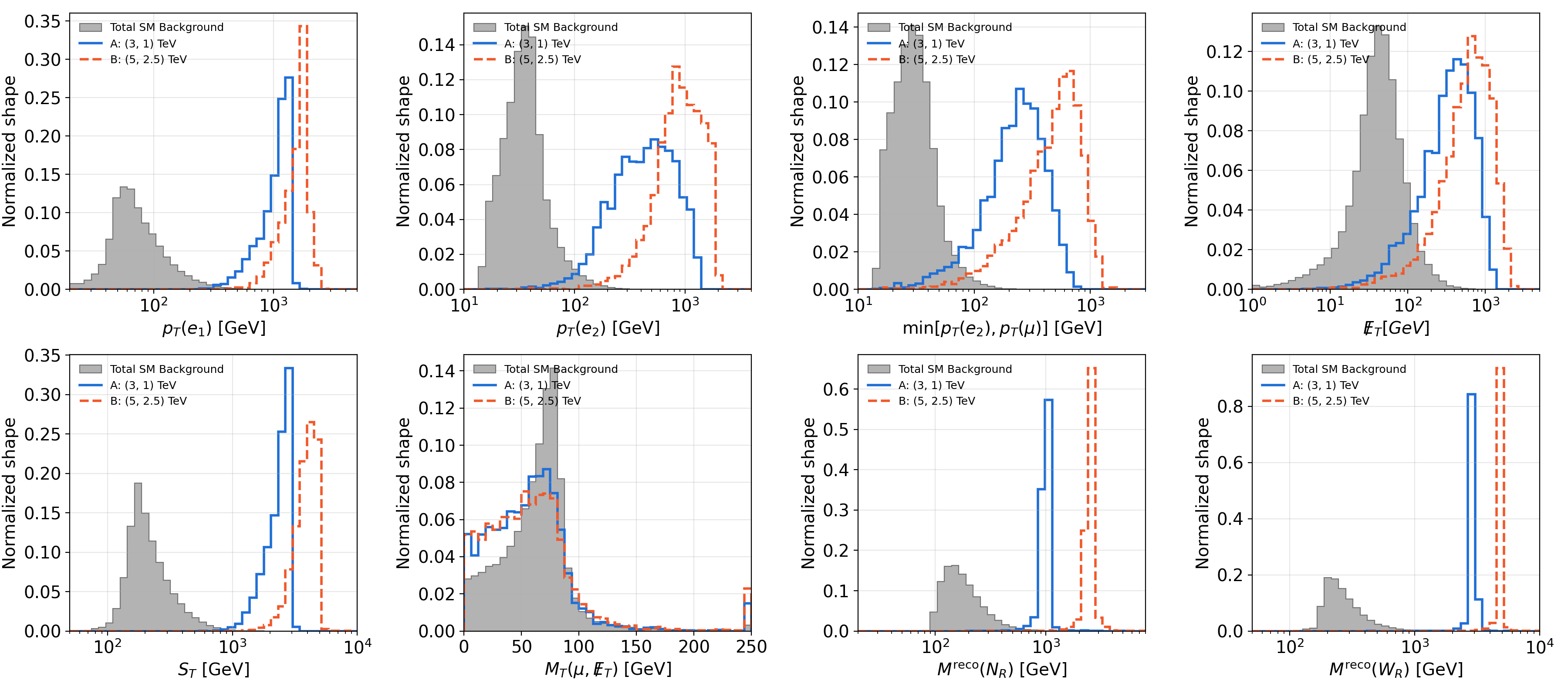}
\caption{\label{fig:dist_eemu} Same as Fig.~\ref{fig:dist_eee}, for the $ee\mu$ channel. The first two panels show the leading and subleading electron, the third shows the softer of the subleading electron and the muon.}
\end{figure*}

\subsection{Event selection}
\label{subsec:selection}
\noindent
We optimise the selection cuts and apply them in the same way to both channels unless otherwise stated. The yields after each step are listed in Tables~\ref{tab:cutflow_eee} and~\ref{tab:cutflow_eemu} for the $eee$ and $ee\mu$ channels, respectively. Below we describe the selection cuts used in our analysis.

\begin{enumerate}
\item \textit{Preselection.} In the $eee$ channel, we require at least three isolated electrons with $|\eta|<2.5$, $p_T>25$~GeV for the leading electron, and $p_T>15$~GeV for the others. We also require no isolated muon with $p_T>10$~GeV and $|\eta|<2.4$. In the $ee\mu$ channel, we require exactly two isolated electrons with the same kinematic requirements as above and exactly one isolated muon with $p_T>15$~GeV and $|\eta|<2.4$. In both channels, the total charge of the three leptons must satisfy $|q_1+q_2+q_3|=1$.

For $Z$+jets and $t\bar t$+jets, we require two prompt leptons and at least one jet with $p_T>30$~GeV and $|\eta|<2.5$. The leading jet takes the place of the missing lepton, with the charge that gives $|\sum Q_i|=1$, and the event is weighted by the number of jets times the fake rate.
The signal loss at this step is dominated by the electron reconstruction and isolation efficiency of the detector simulation, which enters as its third power in the $eee$ channel.
\item \textit{Leading lepton, $p_T^{(1)}>120$~GeV.} The primary lepton from $W_R\to eN_R$ is hard, whereas the leptons of the SM processes peak near $M_{W,Z}/2$. This requirement removes most of the background and practically none of the signal.
\item \textit{Third lepton, $p_T^{(3)}>50$~GeV.} The leptons are ordered in $p_T$ here. The requirement suppresses soft fake leptons, conversions, and unboosted $W$ and $Z$ decays.
\item \textit{Missing transverse momentum, $\slashed{E}_T >100$~GeV.} The neutrino from the boosted $W$ boson carries a large transverse momentum, while the backgrounds without genuine neutrinos, $Z$+jets and $ZZ$, do not.
\item \textit{Scalar sum, $S_T>500$~GeV.} The variable $S_T=\sum_{i=1}^3p_T(\ell_i)+\slashed{E}_T$ tracks the hard scale of the $W_R$.
\item \textit{$Z$ veto.} In the $eee$ channel, the opposite-sign pair closest to the $Z$ mass is used, and in the $ee\mu$ channel, the single $e^+e^-$ pair. The event is rejected if $|M(e^+e^-)-M_Z|<15$~GeV. This removes most of the $WZ$ and $t\bar tZ$ background and keeps practically all of the signal.
\item \textit{$b$-jet veto.} Events with a $b$-tagged jet with $p_T>30$~GeV and $|\eta|<2.5$ are rejected, since the signal contains no heavy-flavour quarks. This suppresses the $t\bar t$+jets and $t\bar tW^\pm$ backgrounds.
\item \textit{Reconstructed $N_R$ mass, $M^{\mathrm{reco}}_{N_R}>200$~GeV.} We use the $N_R$ mass reconstructed as in Sec.~\ref{sec:reconstruction}.
\item \textit{Reconstructed $W_R$ mass, $M^{\mathrm{reco}}_{W_R}>1$~TeV.} The SM background falls steeply in $M^{\mathrm{reco}}_{W_R}$, while the signal peaks at $M_{W_R}$. This requirement removes about three-quarters of the remaining background, without loss of signal for $M_{W_R}$ above about $1$~TeV.
\end{enumerate}

The thresholds of steps 8 and 9 do not depend on the assumed masses, so a single selection serves the whole mass plane. We refer to it as the fixed selection. In Sec.~\ref{subsec:sliding}, we replace these two thresholds with ones that follow the assumed masses.

The fixed selection reduces the background by more than three orders of magnitude relative to the preselection and keeps most of the preselected signal. The remaining background is dominated by $WZ$ production, followed by $t\bar t$+jets and the triple-boson processes.

The shapes of the main kinematic variables after the preselection are compared for the two benchmarks and the total background in Figs.~\ref{fig:dist_eee} and~\ref{fig:dist_eemu}. The lepton transverse momenta, $\slashed{E}_T$, and $S_T$ of the signal are larger than those of the background by about an order of magnitude, and $S_T$ shows the upper edge expected near $M_{W_R}$. The background has a peak at the $Z$ mass in the dielectron mass, which the signal does not have. The reconstructed masses show the largest difference. The signal forms narrow peaks at the true values of $M_{N_R}$ and $M_{W_R}$, while the background is spread over masses below about $1$~TeV. In the fixed selection, these variables enter only through thresholds. The peaks are used more fully by the mass-dependent selection of Sec.~\ref{subsec:sliding}.

\begin{figure*}
\centering
\includegraphics[width=0.46\textwidth]{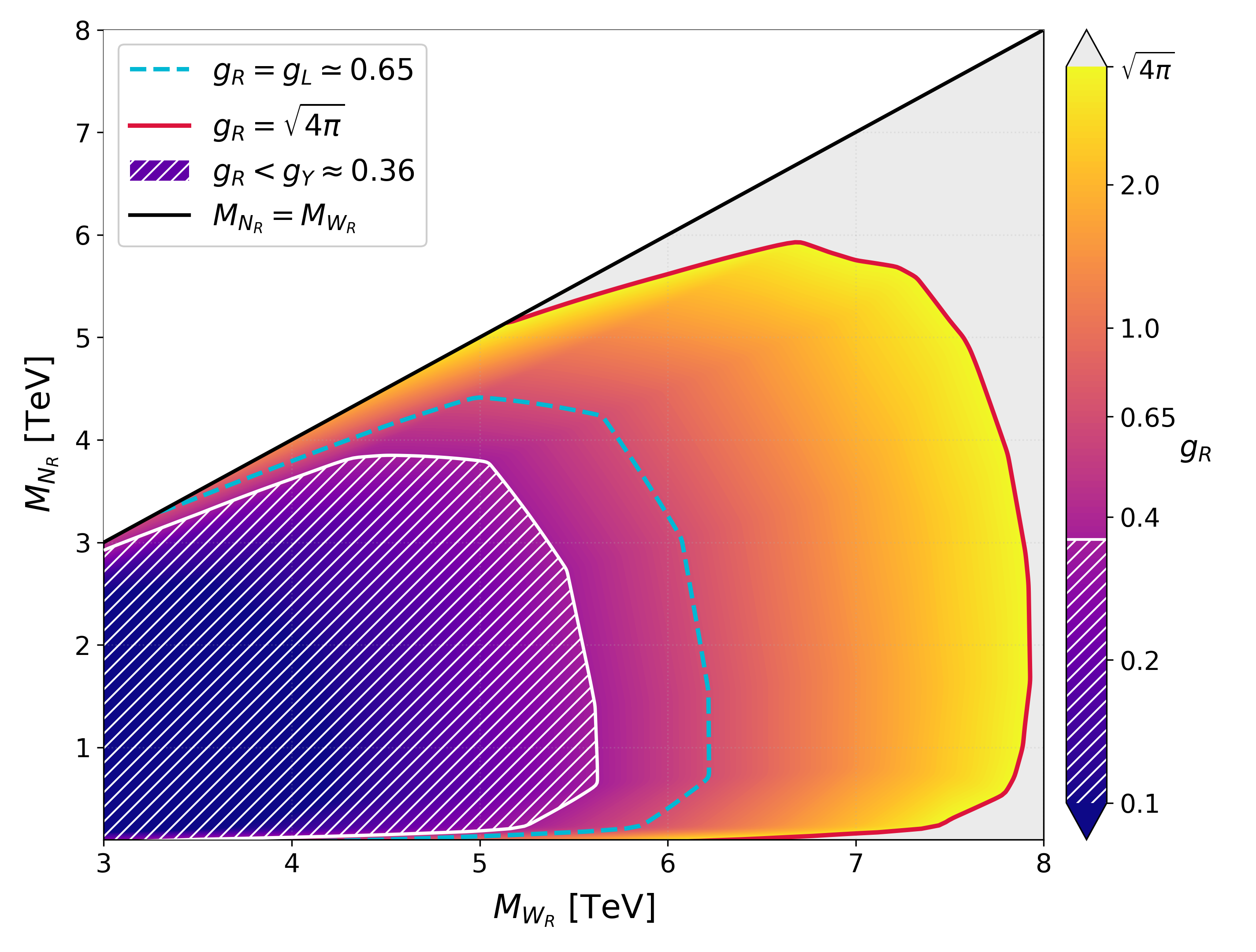}\hfill
\includegraphics[width=0.46\textwidth]{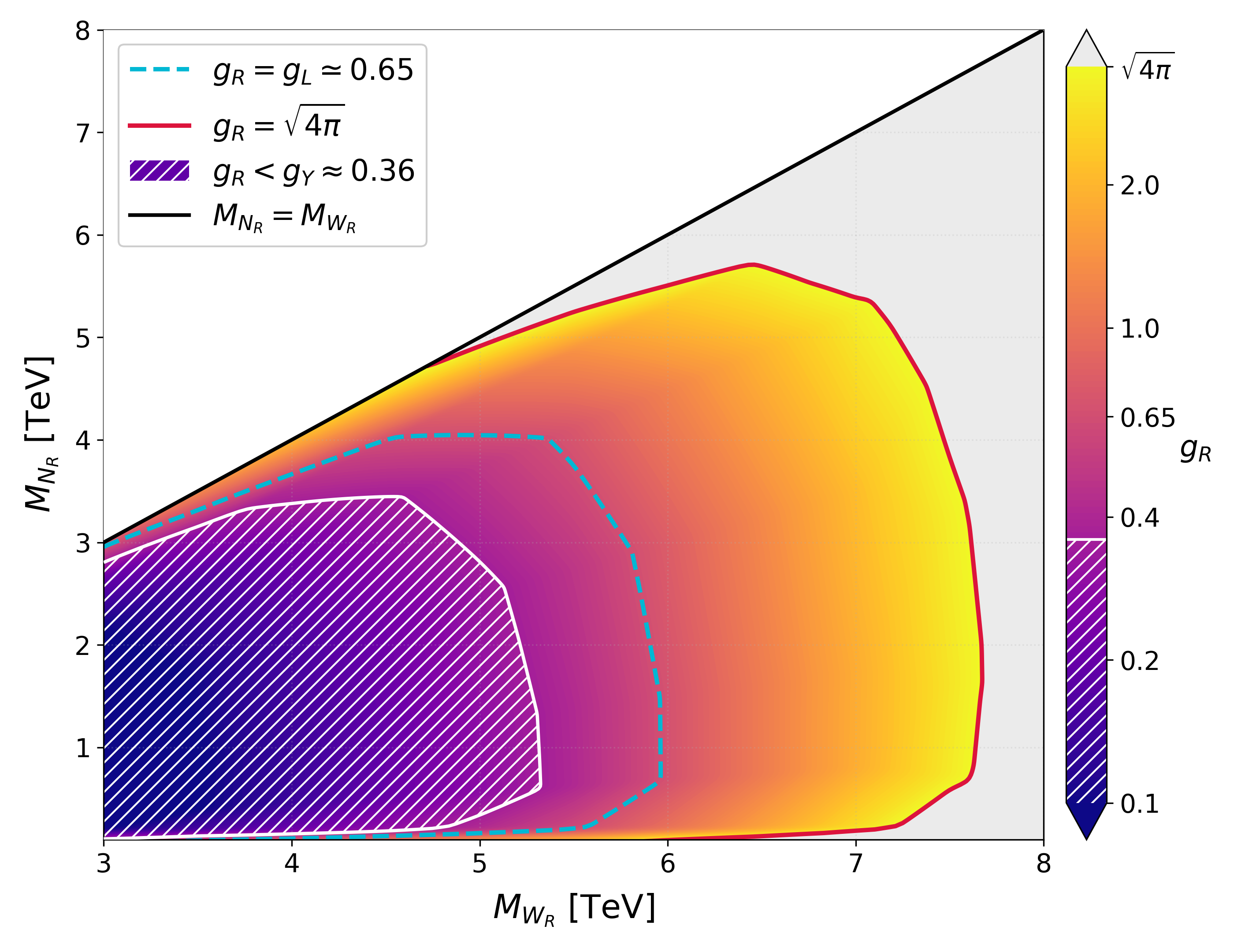}
\caption{Smallest coupling $g_R$ for which the signal can be excluded at $2\sigma$ (left) or discovered at $5\sigma$ (right) at the HL-LHC, in the plane of $M_{W_R}$ and $M_{N_R}$, for the combination of the $eee$ and $ee\mu$ channels with the mass-dependent selection. The colour scale gives $g_R$, with its upper end at the perturbative limit $g_R=\sqrt{4\pi}$. The dashed blue line marks $g_R=g_L\simeq0.65$, the red line the perturbative limit $g_R=\sqrt{4\pi}$, and the black line $M_{N_R}=M_{W_R}$. The hatched region lies below $g_R\approx0.36$, the smallest $g_R$ for which $g_{B-L}$ remains perturbative; mass points there are excluded for any allowed coupling.}
\label{fig:contour_combined}

\includegraphics[width=0.46\textwidth]{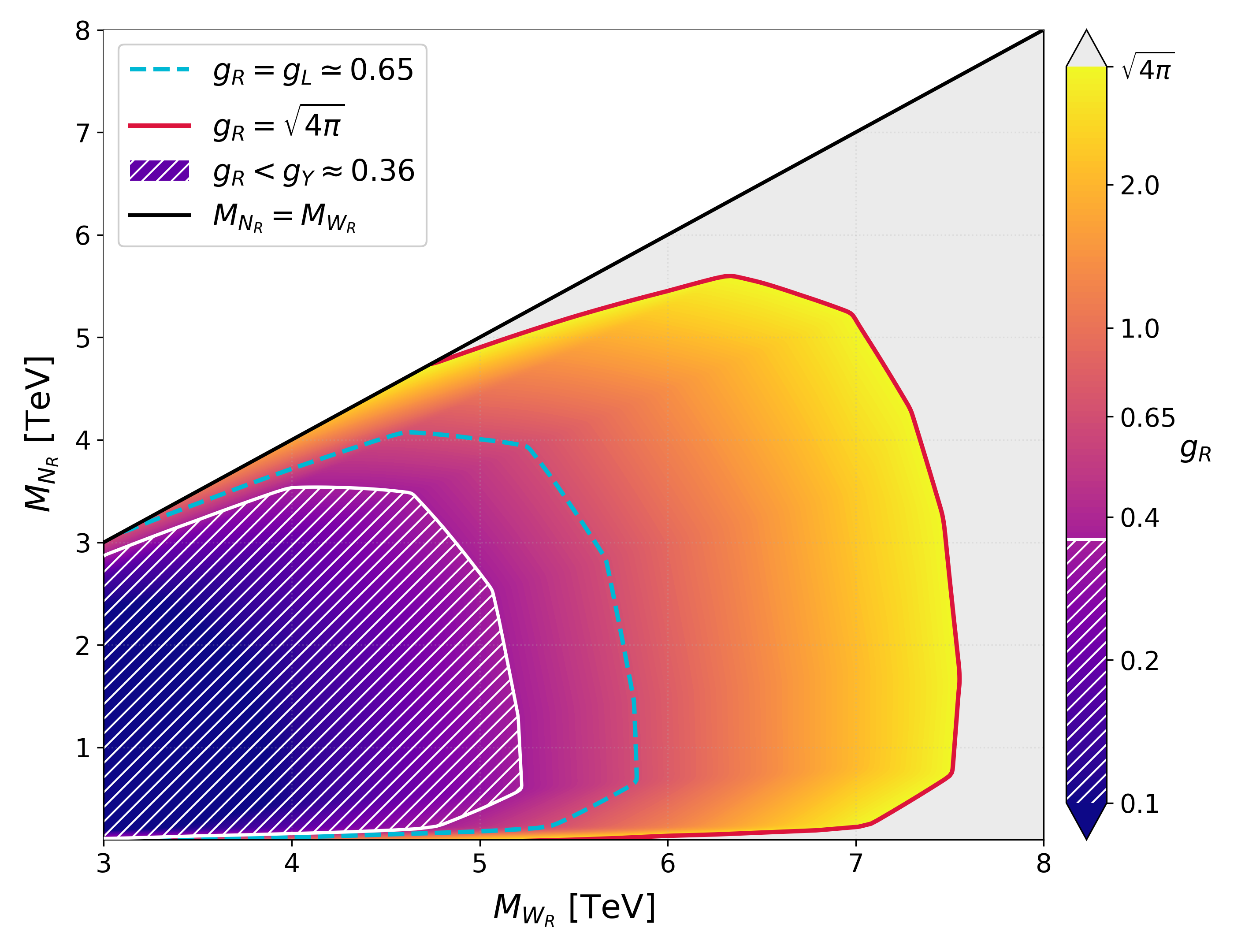}\hfill
\includegraphics[width=0.46\textwidth]{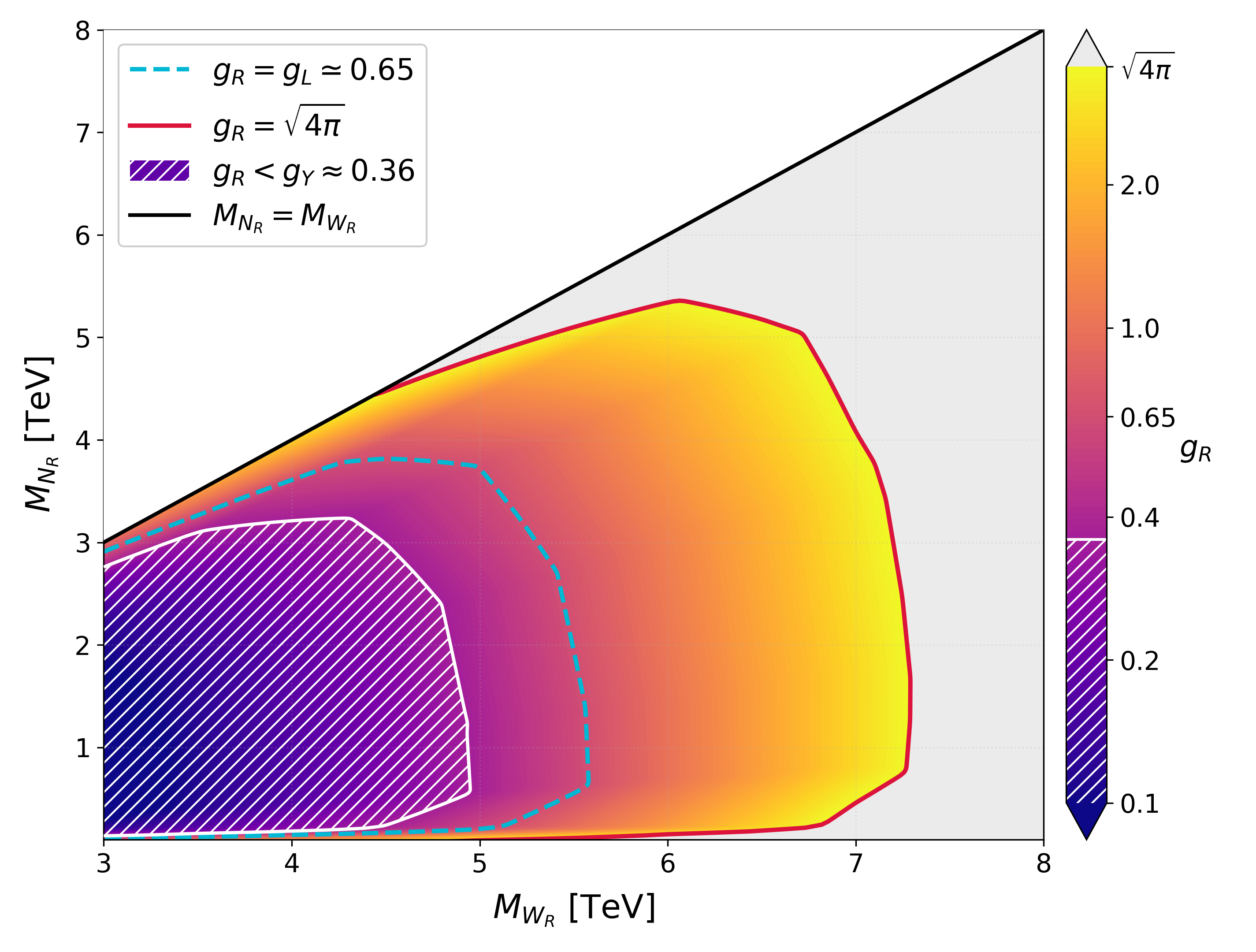}\\[3mm]
\includegraphics[width=0.46\textwidth]{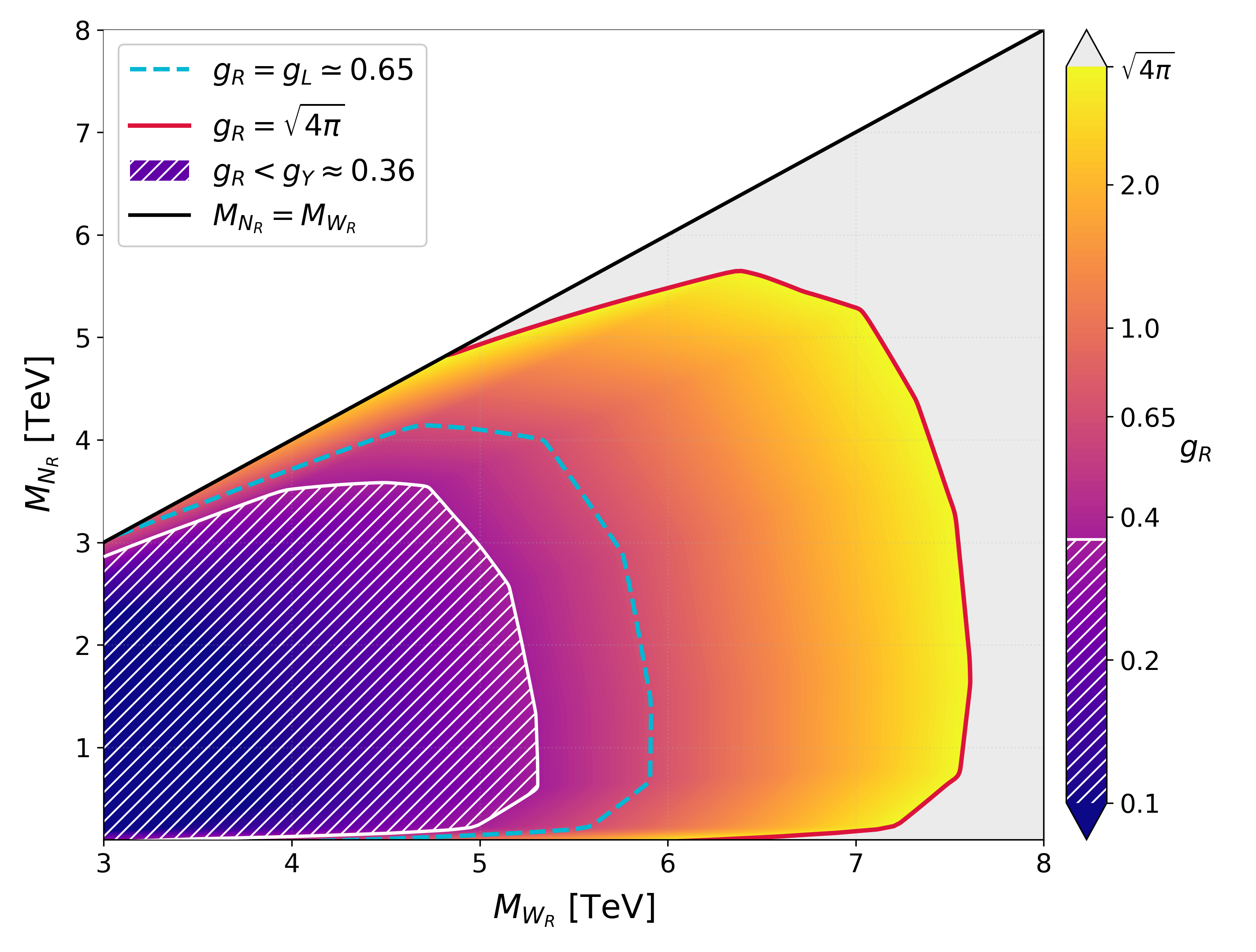}\hfill
\includegraphics[width=0.46\textwidth]{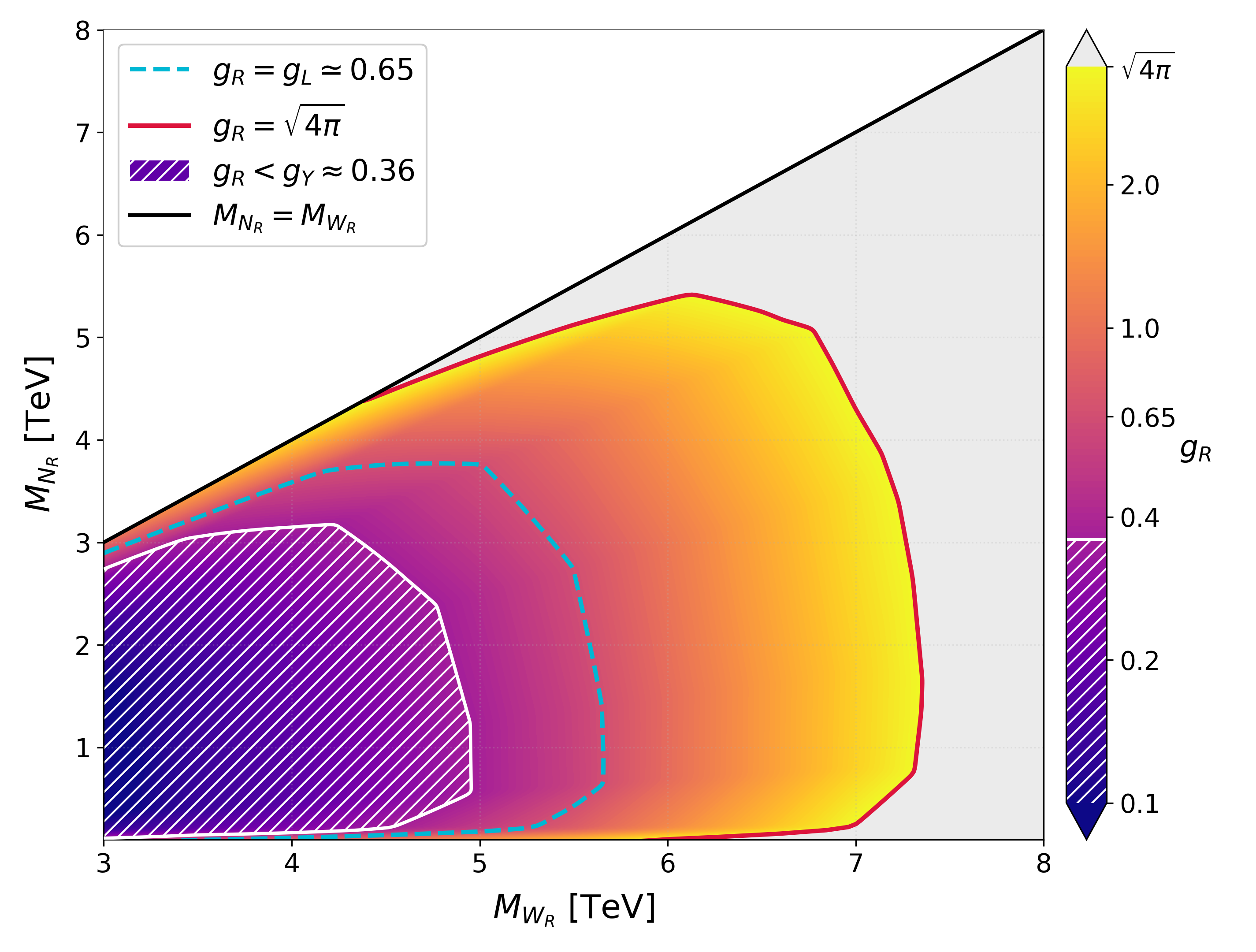}
\caption{Same as Fig.~\ref{fig:contour_combined}, for the $eee$ channel (top) and the $ee\mu$ channel (bottom) separately.}
\label{fig:contour_plot}
\end{figure*}

\section{HL-LHC sensitivity}
\label{sec:results}

\subsection{Statistical treatment}
\label{subsec:statistics}
\noindent
For $\sqrt{s}=14$~TeV and $\mathcal{L}=3~\mathrm{ab}^{-1}$, we compute the expected numbers of signal ($N_S$) and background ($N_B$) events after the selection. For discovery, we use the median significance of a counting experiment with a known background~\cite{Cowan:2010js},
\begin{equation}
\mathcal{Z}_{\rm disc}=\sqrt{2\left[(N_S+N_B)\ln\!\left(1+\frac{N_S}{N_B}\right)-N_S\right]}\,,
\label{eq:significance}
\end{equation}
and for exclusion, the corresponding median significance with which the signal hypothesis is rejected if only background is present~\cite{Cowan:2010js,Bhattiprolu:2020mwi},
\begin{equation}
\mathcal{Z}_{\rm excl}=\sqrt{2\left[N_S-N_B\ln\!\left(1+\frac{N_S}{N_B}\right)\right]}\,.
\label{eq:significance_excl}
\end{equation}
Both reduce to $N_S/\sqrt{N_B}$ for $N_S\ll N_B$. We use $\mathcal{Z}_{\rm excl}=2$ as the criterion for exclusion and $\mathcal{Z}_{\rm disc}=5$ for discovery. When $N_B$ is well below one event, these asymptotic expressions overstate the significance of a handful of signal events. We therefore also require $N_S\ge3$ for exclusion, the $95\%$ confidence-level upper limit of a Poisson count with no background and no observed events, and $N_S\ge5$ for discovery. In the combination of the two channels, the minimum yields apply to the sum of their signal yields. They matter only for the mass-dependent selection of Sec.~\ref{subsec:sliding}, in the multi-TeV region. Only statistical uncertainties are included here; we discuss systematic uncertainties in Sec.~\ref{sec:discussion}.

The two channels are statistically independent, since the preselection makes them orthogonal. Their combination is a two-bin counting experiment, whose significance is
\begin{equation}
\mathcal{Z}_{\rm comb}^2=\mathcal{Z}^2(N_S^{eee},N_B^{eee})+\mathcal{Z}^2(N_S^{ee\mu},N_B^{ee\mu})\,,
\label{eq:combination}
\end{equation}
for both the discovery and the exclusion significance. Without systematic uncertainties, this is exact for the Asimov data set, since the logarithms of the likelihood ratios of independent bins add. Merging the two channels into a single count gives almost the same result, because their signal-to-background ratios are similar.

Since $N_S$ scales as $g_R^2$, the smallest coupling for which a mass point reaches a given significance is
\begin{equation}
g_R^{\rm min}=g_R^{\rm ref}\sqrt{\mu^{\rm thr}}\,,
\label{eq:gmin}
\end{equation}
where $\mu^{\rm thr}$ is the factor by which the signal yields at the reference coupling $g_R^{\rm ref}=0.65$ must be scaled to reach the significance threshold, including the minimum yields above. For a single channel and the fixed selection, the background is the same for all mass points, so a single threshold yield $N_S^{\rm thr}$ applies, and Eq.~\eqref{eq:gmin} reduces to $g_R^{\rm min}=g_R^{\rm ref}\sqrt{N_S^{\rm thr}/N_S(g_R^{\rm ref})}$.

\subsection{Mass-dependent selection}
\label{subsec:sliding}
\noindent
The fixed selection is the same for all masses. For a multi-TeV $W_R$, it keeps a background of several tens of events, mostly $WZ$ at reconstructed masses of $1$ to $2$~TeV, which the signal does not populate. For each mass hypothesis $(M_{W_R},M_{N_R})$, we therefore replace steps 8 and 9 of Sec.~\ref{subsec:selection} with
\begin{align}
M^{\mathrm{reco}}_{N_R}&>\max\!\big(200~\mathrm{GeV},\,f_N\,M_{N_R}\big)\,,\nonumber\\
M^{\mathrm{reco}}_{W_R}&>\max\!\big(1~\mathrm{TeV},\,f_W\,M_{W_R}\big)\,,
\label{eq:sliding}
\end{align}
with $f_N=0.5$ and $f_W=0.7$. Because the reconstructed masses peak at the true values (Sec.~\ref{subsec:performance}), these requirements keep almost all of the signal, with the largest loss, of up to about $10\%$, for $M_{N_R}$ close to $M_{W_R}$. The exclusion reach depends little on the precise fractions, and the discovery reach is stable for $f_W\ge0.7$.

For $M_{W_R}\gtrsim3$~TeV, the thresholds of Eq.~\eqref{eq:sliding} lie beyond the range populated by our background samples, and very few simulated background events survive. We fit the distributions of $M^{\mathrm{reco}}_{N_R}$ and $M^{\mathrm{reco}}_{W_R}$ of the background that passes the fixed selection with exponential functions, starting at three different points of each distribution ($0.2$, $0.4$, and $0.6$~TeV for $M^{\mathrm{reco}}_{N_R}$, and $1$, $1.25$, and $1.5$~TeV for $M^{\mathrm{reco}}_{W_R}$) and taking the largest extrapolation for each variable. Each fit ignores the requirement on the other variable and is therefore an upper estimate of the background. We use the smaller of the two estimates, or the simulated background if that is larger.

\begin{table*}
\caption{\label{tab:mass_reach} Reach in $M_{W_R}$ in TeV, for $g_R=g_L\simeq0.65$ at $2\sigma$ (exclusion) and $5\sigma$ (discovery), as a function of $M_{N_R}$, for the fixed selection of Sec.~\ref{subsec:selection} and the mass-dependent selection of Sec.~\ref{subsec:sliding}. The values are interpolated on the grid of signal points and rounded to $0.1$~TeV.}
{\renewcommand{\arraystretch}{1.1}\setlength{\tabcolsep}{1.5pt}\small
\begin{tabular*}{\textwidth}{@{\extracolsep{\fill}}crrrrrrrrrrrrr}
\hline\hline
& \multicolumn{6}{c}{Fixed selection} & & \multicolumn{6}{c}{Mass-dependent selection} \\
\cmidrule(lr){2-7}\cmidrule(lr){9-14}
$M_{N_R}$ & \multicolumn{2}{c}{$eee$} & \multicolumn{2}{c}{$ee\mu$} & \multicolumn{2}{c}{Combined} & & \multicolumn{2}{c}{$eee$} & \multicolumn{2}{c}{$ee\mu$} & \multicolumn{2}{c}{Combined} \\
\cmidrule(lr){2-3}\cmidrule(lr){4-5}\cmidrule(lr){6-7}\cmidrule(lr){9-10}\cmidrule(lr){11-12}\cmidrule(lr){13-14}
(TeV) & $2\sigma$ & $5\sigma$ & $2\sigma$ & $5\sigma$ & $2\sigma$ & $5\sigma$ & & $2\sigma$ & $5\sigma$ & $2\sigma$ & $5\sigma$ & $2\sigma$ & $5\sigma$ \\
\hline
0.2 & 4.6 & 4.1 & 4.7 & 4.3 & 4.9 & 4.4 & & 5.3 & 5.1 & 5.6 & 5.3 & 5.8 & 5.6 \\
0.5 & 5.0 & 4.6 & 5.0 & 4.6 & 5.2 & 4.8 & & 5.7 & 5.4 & 5.8 & 5.6 & 6.1 & 5.8 \\
1.0 & 5.1 & 4.6 & 5.1 & 4.6 & 5.3 & 4.8 & & 5.8 & 5.6 & 5.9 & 5.7 & 6.2 & 6.0 \\
1.5 & 5.1 & 4.6 & 5.1 & 4.6 & 5.3 & 4.8 & & 5.8 & 5.6 & 5.9 & 5.6 & 6.2 & 6.0 \\
2.5 & 5.0 & 4.4 & 4.9 & 4.3 & 5.2 & 4.6 & & 5.7 & 5.4 & 5.8 & 5.5 & 6.1 & 5.9 \\
\hline\hline
\end{tabular*}}
\end{table*}

\subsection{Reach in the $(M_{W_R},M_{N_R})$ plane}
\label{subsec:reach}
\noindent
Figure~\ref{fig:contour_combined} shows the smallest coupling for which the combination of the two channels reaches $\mathcal{Z}_{\rm excl}=2$ (left) or $\mathcal{Z}_{\rm disc}=5$ (right) in the plane of $M_{W_R}$ and $M_{N_R}$ with the mass-dependent selection. Figure~\ref{fig:contour_plot} shows the same for the two channels separately. The values are obtained from the signal yields using Eq.~\eqref{eq:gmin}, and points outside the red line would require couplings beyond the perturbative limit. Several features can be read from the maps.

The sensitivity depends strongly on $M_{W_R}$. The smallest coupling that can be excluded grows steeply with $M_{W_R}$, following the fall of the production cross section, from below the theoretical minimum $g_R \approx 0.36$ for  $M_{W_R}\lesssim5$~TeV, where every allowed coupling is excluded, to the SM-like value near $6$~TeV. The single channels require somewhat larger couplings than the combination, increasingly so at high mass.

At fixed coupling, the reach is nearly independent of $M_{N_R}$ over a wide range. Table~\ref{tab:mass_reach} lists the reach in $M_{W_R}$ for $g_R=g_L\simeq0.65$ at five values of $M_{N_R}$, for the fixed and mass-dependent selections. For $0.5\le M_{N_R}\le2.5$~TeV, the combined reach of the mass-dependent selection varies by at most $0.2$~TeV. The sensitivity weakens slowly towards the lowest $M_{N_R}$, and when $M_{N_R}$ approaches $M_{W_R}$, where the decay $W_R\to eN_R$ is suppressed by phase space and the contours of constant coupling close towards the diagonal.

The two channels have similar sensitivity. With the fixed selection, the smaller background of the $eee$ channel compensates for its smaller signal yield. With the mass-dependent selection, where the background at high mass is small in both channels and the reach is limited by the number of signal events, the $ee\mu$ channel reaches slightly further because of its larger signal yield. The combination extends the reach by $0.2$ to $0.4$~TeV relative to the better single channel.

The mass-dependent selection extends the combined reach for $g_R=0.65$ from $5.3$ to $6.2$~TeV for exclusion and from $4.8$ to $6.0$~TeV for discovery, at the values of $M_{N_R}$ where the reach is largest. The gain arises because, for $M_{W_R}\gtrsim4.5$~TeV, the background of the mass-dependent selection is below one event, so the reach is set by the number of signal events. For other values of the coupling, the reach follows from the maps. For $g_R\gtrsim1.5$, the combined exclusion extends to the upper end of our grid at $8$~TeV. These large couplings are, however, beyond the range in which the narrow-width approximation is valid, as discussed in Sec.~\ref{sec:discussion}.

\section{Discussion}
\label{sec:discussion}

\subsection{Comparison with existing searches}
\noindent
The Run~2 $\ell\ell jj$ searches of CMS~\cite{CMS:2021dzb} and ATLAS~\cite{ATLAS:2023cjo} exclude $W_R$ masses up to $4.7$ to $6.4$~TeV for $g_R=g_L$, depending on the flavour channel and $M_{N_R}$. These limits are derived for $N_R\to\ell jj$ through a virtual $W_R$, with a branching ratio close to one. In the model studied here, the heavy neutrino decays through its mixing, and the $\ell\ell jj$ final state arises only from $N_R\to\ell W\to\ell jj$, with $\mathrm{BR}(N_R\to\ell W)\,\mathrm{BR}(W\to jj)\approx0.34$. Reinterpreting the published limits with the $\sigma\times\mathrm{BR}$ of this model gives lower bounds of about $4.1$~TeV from CMS and about $5$~TeV from ATLAS for $M_{N_R}=M_{W_R}/2$. The combined exclusion reach of about $6.2$~TeV found here with the mass-dependent selection therefore extends well beyond the present bounds on this scenario. It is also comparable to the reach of about $6$~TeV found at the HL-LHC for the boosted $W$ channel~\cite{ThomasArun:2021rwf}. The $\ell\ell jj$ searches will themselves extend their reach with the HL-LHC data, so the trilepton channel is not expected to set the strongest limit on its own. We nevertheless consider it worth pursuing for two reasons. First, it is an independent measurement. Its backgrounds, dominated by $WZ$ production, and the corresponding systematic uncertainties differ from those of the final states with jets, where the modelling of multijet and top-quark backgrounds at high $p_T$ is the main difficulty. Second, if a signal is seen in a channel with jets, the presence or absence of a trilepton signal distinguishes mixing-induced decays from the KS process, and it gives measurements of $M_{N_R}$ and $M_{W_R}$ that do not depend on the reconstruction of jets.

\subsection{Signal modelling and assumptions}
\noindent
Our results rely on a few assumptions about the signal. The scaling of the yield with $g_R^2$ and Eq.~\eqref{eq:gmin} assume a narrow $W_R$. Above $g_R\simeq1.5$, the narrow-width approximation is no longer reliable, and the off-shell production of the $W_R$, which is enhanced at high masses by the parton luminosities, would move part of the signal to lower reconstructed masses, where the mass-dependent selection rejects it. Thus, the upper part of the colour scale should be read as indicative. The signal $K$ factor of $1.3$ is taken to be independent of the masses. We also assumed that $N_R$ decays through its mixing with the electron neutrino, with the branching ratios of Sec.~\ref{sec:model}. If the mixing is so small that the three-body decay through a virtual $W_R$ competes, the signal is diluted in favour of the $\ell\ell jj$ final state. Mixing with the muon or tau neutrino would populate other flavour combinations, to which the same reconstruction can be applied with the roles of the leptons adapted. Neither case was studied here. Finally, we treated $\mu_S$ as small enough to neglect lepton-number-violating events. If $\Delta M$ were comparable to $\Gamma_N$, a fraction $R_{\ell\ell}/(1+R_{\ell\ell})$ of the signal events would have the charge pattern of the Majorana case and would be assigned incorrectly by the procedure of Sec.~\ref{subsec:assignment}.
\subsection{Background modelling and systematic uncertainties}
\noindent
The background in the multi-TeV region is the least well-controlled part of this study. For $M_{W_R}\gtrsim3$~TeV, it is obtained from the exponential extrapolation of Sec.~\ref{subsec:sliding} rather than from simulated events. If the extrapolated background were ten times larger, the combined exclusion reach for $g_R=0.65$ would decrease by about $0.1$~TeV and the discovery reach by about $0.3$~TeV. The exclusion reach is robust because it is limited by the minimum signal yield, while the discovery reach depends more on the background. A dedicated simulation of the background tails, with generator-level requirements on the invariant mass of the leptons and neutrinos, would replace the extrapolation, and should include $W\gamma$ and $Z\gamma$ production with photon conversions and the misidentification of the electron charge, which our detector simulation does not model.

The reducible backgrounds rely on a constant jet fake rate of $10^{-4}$ for electrons and muons, whereas the actual rates depend on the $p_T$, the pseudorapidity, and the flavour of the jet and are determined from data. In our model, the fake lepton carries the full momentum of the jet, which places these events at higher reconstructed masses than a realistic fake lepton would. A fake rate three times larger would increase the background of the fixed selection by about $20\%$ and weaken the coupling reach by only a few per cent. The $K$ factors of Table~\ref{tab:kfactors} are inclusive. At the high invariant masses relevant here, electroweak corrections reduce the $WZ$ cross section, so this choice is conservative. A real analysis would estimate these backgrounds with data-driven methods.

The significances in this paper are statistical only. To gauge the effect of systematic uncertainties, we added a relative uncertainty of $10\%$ on the background of each channel, uncorrelated between the channels, using the extensions of Eqs.~\eqref{eq:significance} and~\eqref{eq:significance_excl} given in Refs.~\cite{Cowan:2010js,Bhattiprolu:2020mwi}. For the fixed selection, the combined reach for $g_R=0.65$ then decreases by about $0.1$~TeV for both exclusion and discovery. The effect is moderate because the fixed selection leaves only a few tens of background events, whose statistical uncertainty is already comparable to the assumed systematic one. For the mass-dependent selection, the background in the multi-TeV region is below one event, and the reach does not change. Since the reach in the multi-TeV region is limited by the number of signal events, uncertainties on the signal yield matter more. At $M_{W_R} \approx 5 - 6$~TeV, they arise mainly from the parton distributions at large $x$, from scale variations, and from the constant $K$ factor of $1.3$; we estimate them at $10-20\%$. Each $10\%$ reduction of the signal yield lowers the combined reach for $g_R = 0.65$ by $0.05 - 0.07$~TeV, so the total effect is $0.05 - 0.15$~TeV.

\section{Conclusions}
\label{sec:conclusions}
\noindent
We have studied the production of a $W_R$ boson at the HL-LHC in a left-right symmetric model with an inverse seesaw sector, in the final states $eee$ and $ee\mu$ with missing transverse momentum. The heavy neutrinos of this model are pseudo-Dirac, lepton number is conserved to a good approximation, and the cascade $W_R\to\ell N_R\to\ell\ell W\to\ell\ell\ell\nu$ has a definite charge pattern. This allows the cascade to be reconstructed event by event. Charge conservation fixes the role of each lepton up to a twofold ambiguity in the $eee$ channel and removes the ambiguity in the $ee\mu$ channel, and the $W$-mass constraint fixes the longitudinal momentum of the neutrino up to a second twofold ambiguity. The lepton assignment is almost always correct, the $N_R$ mass is reconstructed with a resolution of a few per cent, and for multi-TeV $W_R$ bosons, the reconstructed $W_R$ mass peaks within about $1\%$ of the true mass.

A selection based on the lepton transverse momenta, $\slashed{E}_T$, $S_T$, vetoes on $Z$ bosons and $b$ jets, and thresholds on the reconstructed $N_R$ and $W_R$ masses reduces the SM background by more than three orders of magnitude relative to the preselection and keeps most of the preselected signal. Replacing the thresholds on the reconstructed masses with ones that follow the assumed masses reduces the background in the multi-TeV region to below one event without loss of signal. For $3~\mathrm{ab}^{-1}$ at $\sqrt{s}=14$~TeV and statistical uncertainties only, the combination of the two channels can exclude a $W_R$ boson with the SM-like coupling $g_R=g_L\simeq0.65$ up to about $6.2$~TeV and discover it up to about $6.0$~TeV for $M_{N_R}$ between $1$ and $1.5$~TeV. The reach varies by at most $0.5$~TeV for $M_{N_R}$ between $0.2$ and $2.5$~TeV. This lies well beyond the Run-2 $\ell\ell jj$ limits reinterpreted for this model, of about $5$~TeV. Because the background of the mass-dependent selection is small, these results are insensitive to a systematic uncertainty on the background, but they depend on the extrapolation of the background into the multi-TeV region. We regard the trilepton channel as complementary to the searches with jets: its backgrounds are independent of theirs, and it allows the heavy neutrino and $W_R$ masses to be measured in a final state that the KS process cannot produce.

The most important steps towards a realistic projection are a dedicated simulation of the high-mass background tails including photon conversions and charge misidentification, a data-driven treatment of the fake-lepton backgrounds, and a simulation with HL-LHC pileup and the Phase~II detector.

\begin{acknowledgments}
\noindent
T.M.\ acknowledges partial support from the SERB/ANRF, Government of India, through the Core Research Grant (CRG) No. CRG/2023/007031. R.S. acknowledges partial support from the Anusandhan National Research Foundation (ANRF), Government of India, through Grant No. CRG/2023/008234.\medskip 

\noindent
\textbf{AI usage}: The analysis code for this work was developed with the assistance of Claude Code (Anthropic, Claude Opus 5.5), which was also used to help edit parts of the manuscript. The authors reviewed, tested, and validated all AI-assisted code, and checked every statement and reference in the manuscript. The authors take full responsibility for the content of this paper.
\end{acknowledgments}

\appendix

\section{Solution for the neutrino longitudinal momentum}
\label{app:neutrino}
\noindent
Let $\ell_W$ be massless with momentum $p_\ell=(E_\ell,\vec p_T^{\,\ell},p_z^\ell)$, and let the neutrino have $p_\nu=(E_\nu,\vec p_T^{\,\rm miss},p_z^\nu)$ with $E_\nu=\sqrt{(\slashed{E}_T)^2+(p_z^\nu)^2}$. The constraint $(p_\ell+p_\nu)^2=M_W^2$ is equivalent to $2\,p_\ell\cdot p_\nu=M_W^2$, that is,
\begin{equation}
E_\ell E_\nu=\frac{a}{2}+p_z^\ell\,p_z^\nu\,,\qquad a=M_W^2+2\,\vec p_T^{\,\ell}\!\cdot\vec p_T^{\,\rm miss}\,.
\end{equation}

Squaring this relation and using $E_\ell^2-(p_z^\ell)^2=p_T^2(\ell)$ gives the quadratic equation
\begin{equation}
p_T^2(\ell)\,(p_z^\nu)^2-a\,p_z^\ell\,p_z^\nu+E_\ell^2\,(\slashed{E}_T)^2-\frac{a^2}{4}=0\,.
\end{equation}
Its discriminant is $a^2E_\ell^2-4\,p_T^2(\ell)\,E_\ell^2\,(\slashed{E}_T)^2$, where we used $(p_z^\ell)^2+p_T^2(\ell)=E_\ell^2$, and the solutions are those of Eq.~\eqref{eq:pz_roots},
\begin{equation}
p_z^{\nu,\pm}=\frac{a\,p_z^\ell\pm E_\ell\sqrt{a^2-4\,p_T^2(\ell)\,(\slashed{E}_T)^2}}{2\,p_T^2(\ell)}\,.
\end{equation}
The discriminant is nonnegative if and only if $a\ge2\,p_T(\ell)\,\slashed{E}_T$, since a negative $a$ cannot satisfy it. With $a=M_W^2+2\,p_T(\ell)\,\slashed{E}_T\cos\Delta\phi$ this is the condition $M_T^2=2\,p_T(\ell)\,\slashed{E}_T\,(1-\cos\Delta\phi)\le M_W^2$ of Eq.~\eqref{eq:mt_condition}. At the boundary, $a=2\,p_T(\ell)\,\slashed{E}_T$, the two roots coincide at $p_z^\nu=\slashed{E}_T\,p_z^\ell/p_T(\ell)=\slashed{E}_T\sinh\eta_\ell$, which corresponds to $\eta_\nu=\eta_\ell$. For smaller values of $a$ the roots are complex, and their real part, $p_z^\nu=a\,p_z^\ell/[2\,p_T^2(\ell)]$, continues this solution to the region $M_T>M_W$.

\bibliography{reference}

\end{document}